\documentclass[review,12pt, number]{elsarticle}
\usepackage{newpxtext}

\usepackage[T1]{fontenc}

\usepackage{threeparttable}
\usepackage{geometry}
\usepackage{soul}
\usepackage{ulem}
\usepackage{float}

\usepackage{amsmath, hyperref, geometry, mathrsfs, amssymb, amsthm}
\usepackage{mathrsfs}
\hypersetup{colorlinks = true}
\usepackage{booktabs}
\usepackage{multirow}
\usepackage{footnote}
\usepackage{mathtools}
\usepackage{enumitem}
\usepackage{nomencl}
\usepackage{changepage}
\makesavenoteenv{tabular}
\usepackage{textcomp}
\usepackage{cleveref}
\usepackage{subcaption}
\usepackage{color}
\usepackage{tikz}
\usepackage[titletoc,toc,title]{appendix}
\crefrangeformat{equation}{Eqs.~(#3#1#4) --~(#5#2#6)}

\usepackage[labelfont=bf]{caption}
\graphicspath{{./picture/}}

\DeclareMathOperator*{\argmaxA}{arg\,max} 
\DeclareMathOperator*{\argminA}{arg\,min} 

\makenomenclature
\usepackage{ifthen}
    \renewcommand{\nomgroup}[1]%
        {\ifthenelse{\equal{#1}{S}}{\item[\textbf{Subscripts}]}{}}

\usepackage{lineno}
\usepackage{makecell}

\journal{Mechanical Systems and Signal Processing}

\begin{document}

\begin{frontmatter}

\title{Bayesian spectral density approach for identification and uncertainty quantification of bridge section's flutter derivatives operated in turbulent flow}

\author[a]{Xiaolei Chu}
\author[a,b]{Wei Cui\texorpdfstring{\corref{correspondingauthor}}}
\ead{cuiwei@tongji.edu.cn}
\author[a]{Peng Liu}
\author[a,b,c]{Lin Zhao}
\author[a,b]{Yaojun Ge}
\address[a]{State Key Lab of Disaster Reduction in Civil Engineering, Tongji University, Shanghai, 200092, China}
\address[b]{Key Laboratory of Transport Industry of Bridge Wind Resistance Technologies, Tongji University, Shanghai, 200092, China}
\address[c]{State Key Laboratory of Mountain Bridge and Tunnel Engineering, Chongqing Jiaotong University, Chongqing, 400074, China}

\cortext[correspondingauthor]{Corresponding author: Wei Cui, Assistant Professor, 207 Wind Engineering Building, Tongji University, 1239 Siping Road, Shanghai, 200092, China.}

\begin{abstract}
This study presents a Bayesian spectral density approach for identification and uncertainty quantification of flutter derivatives of bridge sections utilizing buffeting displacement responses, where the wind tunnel test is conducted in turbulent flow. Different from traditional time-domain approaches (e.g., least square method and stochastic subspace identification), the newly-proposed approach is operated in frequency domain. Based on the affine invariant ensemble sampler algorithm, Markov chain Monte-Carlo sampling is employed to accomplish the Bayesian inference. The probability density function of flutter derivatives is modeled based on complex Wishart distribution, where probability serves as the measure. By the Bayesian spectral density approach, the most probable values and corresponding posterior distributions (namely identification uncertainty here) of each flutter derivative can be obtained at the same time. Firstly, numerical simulations are conducted and the identified results are accurate. Secondly, thin plate model, flutter derivatives of which have theoretical solutions, is chosen to be tested in turbulent flow for the sake of verification. The identified results of thin plate model are consistent with the theoretical solutions. Thirdly, the center-slotted girder model, which is widely-utilized long-span bridge sections in current engineering practice, is employed to investigate the applicability of the proposed approach on a general bridge section. For the center-slotted girder model, the flutter derivatives are also extracted by least square method in uniform flow to cross validate the newly-proposed approach. The identified results by two different approaches are compatible.

\end{abstract}
\end{frontmatter}
\section*{Highlights}
\begin{itemize}
  \item Flutter derivatives is identified in frequency domain and in turbulent flow
  \item Flutter derivatives are probabilistically modeled by complex Wishart distribution
  \item Markov chain Monte-Carlo sampling is employed for uncertainty quantification
  \item Applicability of the proposed approach is verified by wind tunnel tests
\end{itemize}

\newpage


\section{Introduction} 
\label{sec:introduction}
Flutter derivatives (FDs) are of vital importance for long-span bridges, which are employed to estimate the flutter critical wind speed \cite{ji2020probabilistic,chu2021probabilistic} and buffeting responses \cite{simiu1996wind}. Conventionally, there are two popular identification schemes to extract FDs in the wind tunnel test. The first one is achieved by a forced vibration test, where the bridge sectional model is assumed to be rigid and supported by a machine (which can move as the predefined displacement's time series) \cite{matsumoto1996aerodynamic,siedziako2018enhanced}. The second one is by free vibration test, where the bridge sectional model is suspended by springs to move in vertical, torsional, and even lateral directions \cite{scanlan1971air,sarkar1992system}. Two schemes mentioned above are both time-domain methods.

As for the forced vibration test, FDs are determined by moving the bridge sectional model in a predefined motion and the aerodynamic forces are being measured at the same time \cite{diana2004forced,sarkar2009comparative}. Usually, forced vibration tests, involved with sinusoidal vertical, torsional, and horizontal motions, are conducted for designs of bridges \cite{cao2012identification}. Recently, Zhao et al. \cite{zhao2020novel} and Siedziako et al. \cite{siedziako2017enhanced} developed the new equipments, which can simultaneously make the bridge sectional model move randomly in vertical, torsional, and lateral directions. As for the free vibration test, there are lots of identification methodologies developed to extract FDs in uniform flow, such as the unifying least-squares method \cite{gu2000identification,bartoli2009toward} and iterative least-squares method \cite{chowdhury2005experimental}. Several researchers also identify the lateral-motion related FDs \cite{chen2002identification,xu2012determination}. Furthermore, some stochastic identification methodologies are developed to extract FDs by utilizing buffeting displacement responses. Qin and Gu \cite{gu2004direct,qin2004determination}, Janesupasaeree and Boonyapinyo \cite{janesupasaeree2011determination}, and Mishra et al. \cite{mishra2006identification} developed the covariance-driven stochastic subspace identification algorithm (SSI) to extract FDs in turbulent flow. Boonyapinyo and Janesupasaeree \cite{boonyapinyo2010data} also developed the data-driven SSI algorithm to extract FDs in turbulent flow. Recently, Wu et al. \cite{wu2021identification} utilized the unscented Kalman filter approach for identification of linear and nonlinear FDs of bridge decks, which can be operated either by free vibration or by buffeting response. However, all methodologies listed above are operated in time domain. Identification of FDs in frequency domain is rarely explored. Furthermore, substantial emphasis has been laid on FDs' experimental uncertainty (i.e., conduct the experiments under the same preset condition many times to get a set of different identified FDs, then get the probability density function of FDs) \cite{seo2011estimation,mannini2015aerodynamic}. But the randomness of FDs in each single wind tunnel test, namely identification uncertainty in this study, is also rarely discussed from a probabilistic perspective. As a result, the purpose of this study is to establish a probabilistic model to analyze the probability density functions (PDFs) of FDs in each wind tunnel test.

Except for SSI and Kalman filter, Bayesian system identification approach is another popular stochastic identification methodology for parameter inference of predefined mathematical models using input-output or output-only dynamic measurements, where probability serves as the measure \cite{beck2010bayesian}. Beck \cite{beck2010bayesian} detailedly illustrated the probability logic with Bayesian updating, providing a rigorous framework to quantify modeling uncertainty and to perform system identification \cite{huang2019state}. Bayesian system identification has been successfully applied on the fields of structural modal identification \cite{au2011fast,zhang2016fast,ni2016fast,au2016fundamental,zhang2016fundamental,au2012ambient}, damage detection \cite{sohn1997bayesian,huang2017hierarchical,zhang2021efficient,wang2021sparse}, and identification of soil stratification \cite{wang2013probabilistic,cao2019bayesian}, etc. Besides, Yuen \cite{yuen2010bayesian} employed the complex Wishart distribution \cite{goodman1963distribution} (namely Bayesian spectral density approach in \cite{yuen2010bayesian}) to infer the most probable values (MPVs) and corresponding posterior uncertainty of structural parameters (modal frequencies, damping ratios, mode shapes) utilizing the power spectrum density (PSD) of the random dynamic responses, which was a paradigm for parameter inference in a stochastic process. The identified quality depends on the accuracy of the predefined mathematical model \cite{yuen2010bayesian}. Au \cite{au2017operational} developed the fast Bayesian FFT algorithm, where the non-overlapping average is avoided compared with the Bayesian spectral density approach \cite{yuen2010bayesian}. Au \cite{au2016insights} also reveals the apparent mathematical equivalence between the Bayesian FFT method and the Bayesian spectral density approach. Typically, the posterior PDF of the inferred parameters $\boldsymbol{\theta}$ can be well approximated by a Gaussian distribution $\mathcal{G}(\boldsymbol{\theta}; \hat{\boldsymbol{\theta}},\mathcal{H}(\hat{\boldsymbol{\theta}})^{-1})$ with mean value $\hat{\boldsymbol{\theta}}$ and covariance matrix $\mathcal{H}(\hat{\boldsymbol{\theta}})^{-1}$ \cite{yuen2010bayesian}, where $\mathcal{H}(\hat{\boldsymbol{\theta}})$ denotes the Hessian matrix calculated at $\hat{\boldsymbol{\theta}}$. A more efficient way is to use Markov chain Monte-Carlo (MCMC) sampling, which can obtain the posterior PDF directly \cite{lam2017bayesian,sedehi2020hierarchical,cheung2017new}. In this paper, Bayesian spectral density approach \cite{yuen2010bayesian} is employed to extract FDs in turbulent flow utilizing buffeting displacement responses, where the affine invariant ensemble sampler (AIES) MCMC \cite{marelli2014uqlab,goodman2010ensemble} is used to obtain the posterior PDFs of FDs.

\section{Bayesian spectral density formulation and uncertainty quantification}
\subsection{Spectral density characteristics of bridge sectional model}
\label{sec:spectral_density_formulation}
\begin{figure}[H]
    \centering
    \includegraphics[]{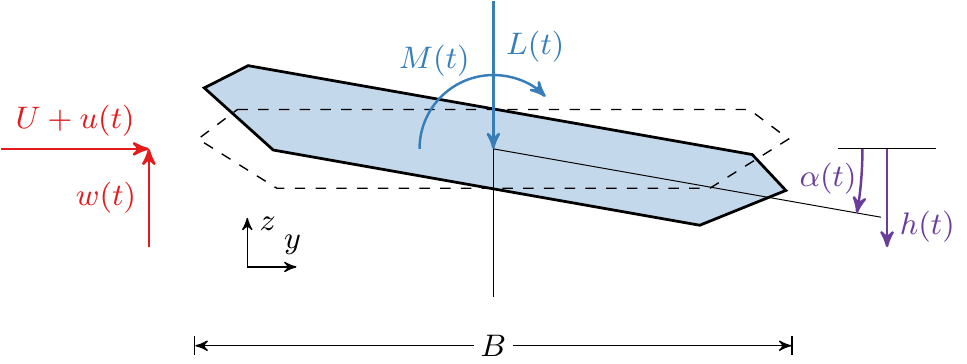}
    \caption{Bridge section and coordinate definition}
	\label{fig:bridge_section}
\end{figure}
In the wind tunnel test, a bridge sectional model with two degree-of-freedom (DOF), i.e., $h$ (vertical motion) and $\alpha$ (torsional motion), is often utilized, as shown in Fig.~\ref{fig:bridge_section}. The dynamic behavior of the sectional model in turbulent flow can be described by the following differential equations \cite{emil1996wind}:
\begin{subequations}
\begin{equation}
\label{eq:2DOF_model}
m\left[\ddot{h}(t)+2 \xi_{h} \omega_{h} \dot{h}(t)+\omega_{h}^{2} h(t)\right]=L_{s e}(t)+L_{b}(t)
\end{equation}
\begin{equation}
I\left[\ddot{\alpha}(t)+2 \xi_{\alpha} \omega_{\alpha} \dot{\alpha}(t)+\omega_{\alpha}^{2} \alpha(t)\right]=M_{s e}(t)+M_{b}(t)
\end{equation}
\end{subequations}
where $m$ and $I$ are the mass and the mass moment of inertia of the deck per unit span, respectively; $\omega_{i}$ is the natural circular frequency, $\xi_{i}$ is the damping ratio ($i=h,\alpha$); $L_{se}$ and $M_{se}$ are the self-excited lift and moment forces, respectively; $L_{b}$ and $M_{b}$ are the buffeting lift and moment forces due to turbulence with zero means. 

The self-excited lift and moment forces are given as follows \cite{simiu1996wind}:
\begin{subequations}
\small
\begin{equation}
L_{s e}=\rho U^{2} B\left[K_{h} H_{1}^{*}\left(K_{h}\right) \frac{\dot{h}}{U}+K_{\alpha} H_{2}^{*}\left(K_{\alpha}\right) \frac{B \dot{\alpha}}{U}+K_{\alpha}^{2} H_{3}^{*}\left(K_{\alpha}\right) \alpha+K_{h}^{2} H_{4}^{*}\left(K_{h}\right) \frac{h}{B}\right]
\end{equation}
\begin{equation}
M_{s e}=\rho U^{2} B^{2}\left[K_{h} A_{1}^{*}\left(K_{h}\right) \frac{\dot{h}}{U}+K_{\alpha} A_{2}^{*}\left(K_{\alpha}\right) \frac{B \dot{\alpha}}{U}+K_{\alpha}^{2} A_{3}^{*}\left(K_{\alpha}\right) \alpha+K_{h}^{2} A_{4}^{*}\left(K_{h}\right) \frac{h}{B}\right]
\end{equation}
\end{subequations}
where $\rho$ is the air mass density; $B$ is the width of the bridge deck; $U$ is the mean wind speed at the bridge deck level; $K_{i} = \omega_{i} B / U$ is the reduced frequency ($i=h,\alpha$); $H_{i}^{*}$ and $A_{i}^{*}$ are the so-called flutter derivatives (FDs), which are usually considered to be related with reduced frequency and cross-sectional shape.

The buffeting lift and moment forces can be defined as \cite{emil1996wind}:
\begin{subequations}
\begin{equation}
\label{eq:fluctuating_wind_speed}
L_{b}(t)=\frac{1}{2} \rho U^{2} B\left[2 C_{L} \frac{u(t)}{U} \chi_{L}(t)+\left(C_{L}^{\prime}+C_{D}\right) \frac{w(t)}{U} \chi_{L}(t)\right]
\end{equation}
\begin{equation}
M_{b}(t)=\frac{1}{2} \rho U^{2} B^{2}\left[2 C_{M} \frac{u(t)}{U} \chi_{M}(t)+\left(C_{M}^{\prime}\right) \frac{w(t)}{U} \chi_{M}(t)\right]
\end{equation}
\end{subequations}
where $C_{L}$, $C_{D}$, and $C_{M}$ are the steady aerodynamic force coefficients; $C_{L}^{'}$ and $C_{M}^{'}$ are the derivatives of $C_{L}$ and $C_{M}$ with respect to the angle of attack, respectively; $u(t)$ and $w(t)$ are the longitudinal and vertical fluctuations of the wind speed, respectively, with zero means; $\chi_{L}$ and $\chi_{M}$ are the lift and moment aerodynamic admittances of the bridge deck.

For the sake of brevity, denote the modified FDs as:
\begin{equation}
\label{eq:modified_FD}
\begin{array}{l}
H_{1}=\frac{\rho B^{2} \omega_{h}}{m} H_{1}^{*}\left(K_{h}\right), 
H_{2}=\frac{\rho B^{3} \omega_{\alpha}}{m} H_{2}^{*}\left(K_{\alpha}\right),
H_{3}=\frac{\rho B^{3} \omega_{\alpha}^{2}}{m} H_{3}^{*}\left(K_{\alpha}\right),
H_{4}=\frac{\rho B^{2} \omega_{h}^{2}}{m} H_{4}^{*}\left(K_{h}\right), \\
A_{1}=\frac{\rho B^{3} \omega_{h}}{I} A_{1}^{*}\left(K_{h}\right),
A_{2}=\frac{\rho B^{4} \omega_{\alpha}}{I} A_{2}^{*}\left(K_{\alpha}\right),
A_{3}=\frac{\rho B^{4} \omega_{\alpha}^{2}}{I} A_{3}^{*}\left(K_{\alpha}\right), 
A_{4}=\frac{\rho B^{3} \omega_{h}^{2}}{I} A_{4}^{*}\left(K_{h}\right)
\end{array}
\end{equation}

Then we have:
\begin{subequations}
\label{eq:2dof_eqa}
\begin{equation}
\ddot{h}(t)+2 \xi_{h} \omega_{h} \dot{h}(t)+\omega_{h}^{2} h(t)=H_{1} \dot{h}(t)+H_{2} \dot{\alpha}(t)+H_{3} \alpha(t)+H_{4} h(t)+\frac{1}{m} L_{b}(t)
\end{equation}
\begin{equation}
\ddot{\alpha}(t)+2 \xi_{\alpha} \omega_{\alpha} \dot{\alpha}(t)+\omega_{\alpha}^{2} \alpha(t)=A_{1} \dot{h}(t)+A_{2} \dot{\alpha}(t)+A_{3} \alpha(t)+A_{4} h(t)+\frac{1}{I} M_{b}(t)
\end{equation}
\end{subequations}

Denote the buffeting displacement response as $\boldsymbol{x}(t) = [h(t), \alpha(t)]^{T}$ and the modified buffeting forces as $\boldsymbol{f}(t) = [\frac{1}{m}L_{b}(t), \frac{1}{I}M_{b}(t)]^{T}$, then Eq.~\eqref{eq:2dof_eqa} becomes:
\begin{equation}
\label{eq:lti_eq}
\begin{array}{c}
\boldsymbol{M} \ddot{\boldsymbol{x}}(t)+\boldsymbol{C} \dot{\boldsymbol{x}}(t)+\boldsymbol{K} \boldsymbol{x}(t)=\boldsymbol{f}(t) \\
\boldsymbol{M}=\left[\begin{array}{ll}
1 & 0 \\
0 & 1
\end{array}\right], \boldsymbol{C}=\left[\begin{array}{cc}
2 \xi_{h} \omega_{h}-H_{1} & -H_{2} \\
-A_{1} & 2 \xi_{\alpha} \omega_{\alpha}-A_{2}
\end{array}\right], \boldsymbol{K}=\left[\begin{array}{cc}
\omega_{h}^{2}-H_{4} & -H_{3} \\
-A_{4} & \omega_{\alpha}^{2}-A_{3}
\end{array}\right]
\end{array}
\end{equation}

Considering that the fluctuations of wind speed $u(t)$ and $w(t)$ in Eq.~\eqref{eq:fluctuating_wind_speed} are random functions of time \cite{qin2004determination}, Eq.~\eqref{eq:lti_eq} means that the bridge sectional model is a time-invariant linear system if the mean wind speed $U$ is given. The frequency response function of the bridge sectional model is:
\begin{equation}
\boldsymbol{\mathrm{H}}(\omega) = \left(\boldsymbol{K} - \omega^{2}\boldsymbol{M} + {\mathrm{i}} \omega \boldsymbol{C} \right)^{-1}
\end{equation}
where $\boldsymbol{K}$, $\boldsymbol{M}$, and $\boldsymbol{C}$ are coupled stiffness matrix, mass matrix, and coupled damping matrix in Eq.~\eqref{eq:lti_eq}, respectively; ${\mathrm{i}}$ is the imaginary unit; $\boldsymbol{\mathrm{H}}(\omega)$ is the frequency response function at the specific circular frequency $\omega$.

The PSD matrix of buffeting displacement response $\boldsymbol{x}$ is:
\begin{equation}
\label{eq:PSD_displacement}
\mathbf{S}_{\boldsymbol{x}}(\omega)=\mathbf{H}(\omega)\mathbf{S}_{\boldsymbol{f}}(\omega)\mathbf{H}^{\star}(\omega)=\mathbf{H}(\omega)\left[\begin{array}{cc}
S_{L}(\omega) & 0 \\
0 & S_{M}(\omega)
\end{array}\right] \mathbf{H}^{\star}(\omega)
\end{equation}
where $(\cdot)^{\star}$ means the conjugate transpose of $(\cdot)$; $\boldsymbol{\mathrm{S}_{f}}(\omega)=\operatorname{diag}\left[\mathrm{S}_{L}(\omega),\mathrm{S}_{M}(\omega)\right]$ means the PSD matrix of the modified buffeting lift force ($\frac{1}{m} L_{b}$) and modified moment force ($\frac{1}{I} M_{b}$) at circular frequency $\omega$, respectively. Theoretically, $S_{L}(\omega)$ and $S_{M}(\omega)$ are not constant in the entire frequency band \cite{jain1996coupled}. Also, the cross-spectrum of buffeting lift force and buffeting moment force is complicated \cite{jain1996coupled}, which is not considered in this study. This study finds that within a narrow frequency band $\mathcal{K} = \{k_{1}\Delta \omega, (k_{1}+1)\Delta \omega, \dots, k_{2} \Delta \omega\}$ ($0<k_{1}<k_{2})$ around the bridge sectional model's resonant circular frequency (i.e., $\omega_{h}$ and $\omega_{\alpha}$), the cross-spectrum of buffeting lift and moment forces can be neglected and $\mathbf{S}_{\boldsymbol{f}}(\mathcal{K})$ can be considered as a constant matrix, which will not influence the identification accuracy. Similar solutions can also be found in Wu et al. \cite{wu2021identification}. Constant matrix $\mathbf{S}_{\boldsymbol{f}}(\mathcal{K})$ can simplify the modeling procedure.

Then we can utilize $\boldsymbol{\mathrm{S}_{x}}(\omega)$ to infer the FDs in the frequency domain. Denote all that we need to identify as the parameter vector $\boldsymbol{\theta}$:
\begin{equation}
\label{eq:definition_theta}
\boldsymbol{\theta} = \left[A_{1}^{\star}, \cdots, A_{4}^{\star}, H_{1}^{\star}, \cdots, H_{4}^{\star}, \mathrm{S}_{L,1}, \mathrm{S}_{M,1}, \cdots, \mathrm{S}_{L,d}, \mathrm{S}_{M,d}\right]^{T}
\end{equation}
where $d=2$ in this study because the bridge sectional model is simplified as 2-DOF in Fig.~\ref{fig:bridge_section} ($d$ will be equal to $3$ if the bridge sectional model is simplified as 3-DOF, i.e., when there are 18 FDs \cite{chowdhury2003new}).

Different from \cite{au2011fast,katafygiotis2001bayesian}, the measurement noise is not considered here because we find that the contribution from measurement noise to the measured response PSDs in the wind tunnel is negligible. It can be proved by Fig.~\ref{fig:plate_PSD_reconstruction} and Fig.~\ref{fig:XHM_PSD_reconstruction}, where the reconstructed PSDs are consistent with the measured response PSDs even though the measurement noise is not considered. Neglecting the measurement noise in this study will not influence the identifying accuracy of FDs. On the other hand, the number of parameters to be inferred can also be reduced by neglecting the measurement noise, which increases the computational efficiency.

\subsection{Spectral density estimator and complex Wishart distribution}
Consider the stochastic vector process $\boldsymbol{x}(t) = \left[ x_{1}(t), x_{2}(t), \dots, x_{d}(t) \right ]^{T}$ and a finite number of discrete data $\boldsymbol{\mathrm{X}}_{N}=\{\boldsymbol{x}(m), m = 0, 1, \dots, N-1 \}$, where $d=2$ in this study due to the 2-DOF bridge sectional model. Based on $\boldsymbol{X}_{N}$, the following discrete estimator of the spectral density matrix of the stochastic process $\boldsymbol{x}(t)$ is introduced \cite{yuen2002spectral}:
\begin{equation}
\label{eq:PSD_Yuen}
\boldsymbol{\mathrm{S}}_{\boldsymbol{x},N}(\omega_{k}) = {\mathscr{X}}_{N}(\omega_{k}) \mathscr{X}_{N}^{*}(\omega_{k}) 
\end{equation}
where $\mathscr{X}_{N}(\omega_{k})$ denotes the (scaled) Fourier Transform of the vector process $\boldsymbol{x}$ at circular frequency $\omega_{k}$, as follows:
\begin{equation}
\mathscr{X}_{N}(\omega_{k}) = \sqrt{\frac{\Delta t}{2 \pi N}} \sum_{m=0}^{N-1}{\boldsymbol{x}(m) \exp^{{-\mathrm{i}} \omega_{k} m \Delta t}}
\end{equation}
where $\omega_{k} = k\Delta \omega$, $k=0,1,\dots,N_{1}-1$ with $N_{1}=\mathrm{INT}(\frac{N+1}{2})$ (denotes the ordinate at the Nyquist frequency), $\Delta \omega = \frac{2 \pi}{T}$, and $T=N\Delta t$.

It is shown that the complex vector $\mathscr{X}_{N}(\omega_{k})$ has a complex multivariate Gaussian distribution with zero mean as $N \rightarrow \infty$ \cite{yuen2002spectral,goodman1963statistical}. Assume now that there is a set of $M$ independent and identically distributed time histories whose realizations correspond to sampling data from the same process $\boldsymbol{x}(t)$ without overlapping. As $N \rightarrow \infty$, the corresponding frequency-domain stochastic processes $\mathscr{X}_{N}^{(m)}(\omega_{k})$, $m = 1, 2, \dots, M$, are independent and follow an identical complex $d$-variate Gaussian distribution with zero mean \cite{goodman1963statistical}. Then, as $N \rightarrow \infty$ and if $M \ge d$, denote the average spectral density estimator as:
\begin{equation}
\label{eq:PSD_sampling}
\boldsymbol{\mathrm{S}}_{N}^{M}(\omega_{k}) = \frac{1}{M} \sum_{m=1}^{M}{\boldsymbol{\mathrm{S}}_{N}^{(m)}(\omega_{k})}
\end{equation}
where $\boldsymbol{\mathrm{S}}_{N}^{M}(\omega_{k})$ follows a central complex Wishart distribution \cite{goodman1963distribution} of dimension $d$ with $M$ sets and the mean value $\mathbb{E}[ \boldsymbol{\mathrm{S}}_{N}^{M}(\omega_{k}) |\boldsymbol{\theta}] = \mathbb{E} [ \boldsymbol{\mathrm{S}}_{N}(\omega_{k}) |\boldsymbol{\theta} ]$ \cite{krishnaiah1976some}:
\begin{equation}
\label{eq:complex_Wishart_distribution}
\begin{aligned}
p\left[\mathbf{S}_{N}^{M}\left(\omega_{k}\right)\right|\boldsymbol{\theta}]= &\frac{\pi^{-d(d-1) / 2} M^{M-d+d^{2}}\left|\mathbf{S}_{N}^{M}\left(\omega_{k}\right)\right|^{M-d}}{\left[\prod_{p=1}^{d}(M-p) !\right] |\mathbb{E}\left[\mathbf{S}_{N}\left(\omega_{k}\right)|\boldsymbol{\theta}\right]|^{M}} \\
& \times \exp \left(-M \cdot \operatorname{tr}\left\{\mathbb{E}\left[\mathbf{S}_{N}\left(\omega_{k}\right)|\boldsymbol{\theta}\right]^{-1} \mathbf{S}_{N}^{M}\left(\omega_{k}\right)\right\}\right)
\end{aligned}
\end{equation}
where the matrix $\mathbb{E} [ \boldsymbol{\mathrm{S}}_{N}(\omega_{k})|\boldsymbol{\theta} ]$ is the theoretical matrix obtained by Eq.~\eqref{eq:PSD_displacement} given the parameter vector $\boldsymbol{\theta}$; the matrix $\mathbf{S}_{N}^{M}(\omega_{k})$ is obtained by Eq.~\eqref{eq:PSD_sampling}, which is the measured PSD matrix by the wind tunnel test; $\left| \cdot \right|$ means the determinant of a matrix; $\operatorname{tr}(\cdot)$ means the trace of a matrix; $p(\cdot)$ means the PDF. Actually, Eq.~\eqref{eq:complex_Wishart_distribution} links the theoretical PSD matrix $\mathbb{E} [ \boldsymbol{\mathrm{S}}_{N}(\omega_{k})|\boldsymbol{\theta} ]$ with the measured PSD matrix $\mathbf{S}_{N}^{M}(\omega_{k})$.

Furthermore, it has been investigated \cite{katafygiotis2001bayesian} that if $N \rightarrow \infty$, the matrices $\mathbf{S}_{N}^{M}(\omega_{k})$ and $\mathbf{S}_{N}^{M}(\omega_{l})$ are independently Wishart distributed for $k \neq l$, that is:
\begin{equation}
\label{eq:pdf_independent}
p[\mathbf{S}_{N}^{M}(\omega_{k}), \mathbf{S}_{N}^{M}(\omega_{l})]=p[\mathbf{S}_{N}^{M}(\omega_{k})] \cdot p[\mathbf{S}_{N}^{M}(\omega_{l})]
\end{equation}

Consider two frequency index sets $\mathcal{K}^{(1)} = \{k_{1}\Delta \omega, (k_{1}+1)\Delta \omega, \dots, k_{2} \Delta \omega\}$ ($k_{1} < k_{2}$) (around $\omega_{h}$) and $\mathcal{K}^{(2)} = \{k_{3}\Delta \omega, (k_{3}+1)\Delta \omega, \dots, k_{4} \Delta \omega\}$ ($k_{2} \le k_{3} < k_{4}$) (around $\omega_{\alpha}$), then we can form the spectral set $\mathbf{S}_{N}^{M,\mathcal{K}^{(1)},\mathcal{K}^{(2)}} = [\mathbf{S}_{N}^{M}(k), k \in \left\{\mathcal{K}^{(1)}, \mathcal{K}^{(2)} \right\}]^{T}$. According to Eq.~\eqref{eq:pdf_independent}, we have:
\begin{equation}
\label{eq:prod_pdf}
p(\mathbf{S}_{N}^{M,\mathcal{K}^{(1)},\mathcal{K}^{(2)}}|\boldsymbol{\theta}) = \prod_{k\in \mathcal{K}^{(1)}}{p[\mathbf{S}_{N}^{M}(k)|\boldsymbol{\theta}]} \cdot \prod_{k\in \mathcal{K}^{(2)}}{p[\mathbf{S}_{N}^{M}(k)|\boldsymbol{\theta}]}
\end{equation}
where $p(\mathbf{S}_{N}^{M,\mathcal{K}^{(1)},\mathcal{K}^{(2)}}|\boldsymbol{\theta})$ is the likelihood function.

\subsection{Identification and uncertainty quantification of flutter derivatives}

The parameter vector $\boldsymbol{\theta}$ is actually a random vector during the process of inference, which is denoted as a random vector $\mathbf{\Theta}$ with known prior distribution $p(\boldsymbol{\theta})$ ($\mathbf{\Theta}\sim p(\boldsymbol{\theta})$) in this paper. Combining the prior $p(\boldsymbol{\theta})$ and the likelihood function $p(\mathbf{S}_{N}^{M,\mathcal{K}^{(1)},\mathcal{K}^{(2)}}|\boldsymbol{\theta})$ in Eq.~\eqref{eq:prod_pdf}, under the context of Bayes' theorem \cite{marelli2014uqlab}, the posterior distribution of $\mathbf{\Theta}$ is established:
\begin{equation}
\label{eq:posterior_prod_pdf}
p(\boldsymbol{\theta}|\boldsymbol{\mathrm{S}}_{N}^{M,\mathcal{K}^{(1)},\mathcal{K}^{(2)}}) = \kappa p(\boldsymbol{\theta})\prod_{k\in \mathcal{K}^{(1)}}{p[\mathbf{S}_{N}^{M}(k)|\boldsymbol{\theta}]} \cdot \prod_{k\in \mathcal{K}^{(2)}}{p[\mathbf{S}_{N}^{M}(k)|\boldsymbol{\theta}]}
\end{equation}
where $\kappa$ is a normalizing constant \cite{yuen2010bayesian}; $p(\boldsymbol{\theta}|\boldsymbol{\mathrm{S}}_{N}^{M,\mathcal{K}^{(1)},\mathcal{K}^{(2)}})$ is the posterior distribution.

By Eq.~\eqref{eq:posterior_prod_pdf}, the most probable value (MPV) $\hat{\boldsymbol{\theta}}$ is:
\begin{equation}
\hat{\boldsymbol{\theta}}=\argmaxA_{\boldsymbol{\theta}} \left[ p(\boldsymbol{\theta}|\boldsymbol{\mathrm{S}}_{N}^{M,\mathcal{K}^{(1)},\mathcal{K}^{(2)}}) \right]
\end{equation}

For individual component $\theta_{i}$ ($i = 1,2,\dots,12$), its marginal PDF can be calculated by integrating over the other components:
\begin{equation}
p(\theta_{i}|\boldsymbol{\mathrm{S}}_{N}^{M,\mathcal{K}^{(1)},\mathcal{K}^{(2)}}))=\int_{\mathbf{\Omega}_{\mathbf{\Theta}_{\sim i}}}{p(\boldsymbol{\theta}|\boldsymbol{\mathrm{S}}_{N}^{M,\mathcal{K}^{(1)},\mathcal{K}^{(2)}}))\mathrm{d}\boldsymbol{\theta}_{\sim i}}
\end{equation}
where $\boldsymbol{\theta}_{\sim i}$ means the parameter vector $\boldsymbol{\theta}$ excluding the $i$-th parameter $\theta_{i}$.

Practically \cite{yuen2010bayesian,au2017operational}, maximizing $p(\boldsymbol{\theta}|\boldsymbol{\mathrm{S}}_{N}^{M,\mathcal{K}^{(1)},\mathcal{K}^{(2)}})$ is equivalent to minimizing the negative log-likelihood function $L(\boldsymbol{\theta})$ of Eq.~\eqref{eq:posterior_prod_pdf}. $L(\boldsymbol{\theta})$ is:
\begin{equation}
\label{eq:NLLF}
\begin{aligned}
L(\boldsymbol{\theta})&=-\ln{p(\boldsymbol{\theta}|\boldsymbol{\mathrm{S}}_{N}^{M,\mathcal{K}^{(1)},\mathcal{K}^{(2)}})} =-\ln \left\{\kappa p(\boldsymbol{\theta}) \prod_{k \in \mathcal{K}^{(1)}} p\left[\mathbf{S}_{N}^{M}(k) \mid \boldsymbol{\theta}\right] \cdot \prod_{k \in \mathcal{K}^{(2)}} p\left[\mathbf{S}_{N}^{M}(k) \mid \boldsymbol{\theta}\right]\right\} \\
&\propto M\left\{\sum_{k=k_{1}}^{k_{2}} \ln \left|\mathbb{E}\left[\mathbf{S}_{N}(k \Delta \omega) \mid \boldsymbol{\theta}\right]\right|+\sum_{k=k_{1}}^{k_{2}} \operatorname{tr}\left[\mathbb{E}\left[\mathbf{S}_{N}(k \Delta \omega) \mid \boldsymbol{\theta}\right]^{-1} \mathbf{S}_{N}^{M}(k \Delta \omega)\right]\right.\\
&\left.+\sum_{k=k_{3}}^{k_{4}} \ln \left|\mathbb{E}\left[\mathbf{S}_{N}(k \Delta \omega) \mid \boldsymbol{\theta}\right]\right|+\sum_{k=k_{3}}^{k_{4}} \operatorname{tr}\left[\mathbb{E}\left[\mathbf{S}_{N}(k \Delta \omega) \mid \boldsymbol{\theta}\right]^{-1} \mathbf{S}_{N}^{M}(k \Delta \omega)\right]\right\} - \ln{p(\boldsymbol{\theta})}
\end{aligned}
\end{equation}

Equivalently, the MPV $\hat{\boldsymbol{\theta}}$ is:
\begin{equation}
\hat{\boldsymbol{\theta}}=\argmaxA_{\boldsymbol{\theta}} \left[ p(\boldsymbol{\theta}|\boldsymbol{\mathrm{S}}_{N}^{M,\mathcal{K}^{(1)},\mathcal{K}^{(2)}}) \right]=\argminA_{\boldsymbol{\theta}} \left[ L(\boldsymbol{\theta}) \right]
\end{equation}

For a globally identifiable model updating problem, the MPV can be estimated by numerically maximizing the posterior PDF, which can be well approximated by a multi-variable Gaussian distribution \cite{yuen2010bayesian}. However, the actual form of the posterior PDF $p(\boldsymbol{\theta}|\boldsymbol{\mathrm{S}}_{N}^{M,\mathcal{K}^{(1)},\mathcal{K}^{(2)}})$ in this study is too complicated to be described as Gaussian distribution. One popular solution is based on Markov chain Monte Carlo (MCMC) sampling \cite{robert2013monte,liu2008monte}. The basic idea of MCMC is to construct stationary chains of samples, and the posterior is the invariant distribution of the Markov chain if the transition probability $\mathcal{\pi} \left( \boldsymbol{\theta}^{(t+1)}|\boldsymbol{\theta}^{(t)} \right)$ fulfills the detailed balance condition
\begin{equation}
\label{eq:detailed_balance}
p(\boldsymbol{\theta}^{(t)}|\boldsymbol{\mathrm{S}}_{N}^{M,\mathcal{K}^{(1)},\mathcal{K}^{(2)}}) \mathcal{\pi}(\boldsymbol{\theta}^{(t+1)}|\boldsymbol{\theta}^{(t)})=p(\boldsymbol{\theta}^{(t+1)}|\boldsymbol{\mathrm{S}}_{N}^{M,\mathcal{K}^{(1)},\mathcal{K}^{(2)}}) \mathcal{\pi}(\boldsymbol{\theta}^{(t)}|\boldsymbol{\theta}^{(t+1)})
\end{equation}

In this study, AIES-based algorithm \cite{goodman2010ensemble} is adopted to better accommodate the potential strong correlation among the parameters within the posterior distribution. The affine invariance property is achieved by generating proposals according to stretch move \cite{goodman2010ensemble}. This refers to proposing a new candidate by
\begin{equation}
\boldsymbol{\theta}_{i}^{(\star)}=\boldsymbol{\theta}_{i}^{(t)}+Z\left(\boldsymbol{\theta}_{j}^{(t)}-\boldsymbol{\theta}_{i}^{(t)}\right), ~ \text{where} ~ Z \sim p(z)=\left\{\begin{array}{ll}
\frac{1}{\sqrt{z}\left(2 \sqrt{a}-\frac{2}{\sqrt{a}}\right)} & \text{if} ~z \in[1 / a, a] \\
0 & \text {otherwise}
\end{array}\right.
\end{equation}

This requires sampling from the distribution $p(z)$ defined by the tuning parameter $a>1$ (practically $a=2$ \cite{marelli2014uqlab,goodman2010ensemble}). The acceptance rate $\alpha\left(\boldsymbol{\theta}_{i}^{(\star)},\boldsymbol{\theta}_{i}^{(t)} \right)$ is
\begin{equation}
\alpha\left(\boldsymbol{\theta}_{i}^{(\star)},\boldsymbol{\theta}_{i}^{(t)} \right) = \operatorname{min}\left \{1, z^{n-1} \frac{p(\boldsymbol{\theta}_{i}^{(\star)}|\boldsymbol{\mathrm{S}}_{N}^{M,\mathcal{K}^{(1)},\mathcal{K}^{(2)}})}{p(\boldsymbol{\theta}_{i}^{(t)}|\boldsymbol{\mathrm{S}}_{N}^{M,\mathcal{K}^{(1)},\mathcal{K}^{(2)}})} \right \}
\end{equation}
where $n$ is the dimension of the parameter space. 

The convergence criteria for the posterior distributions of all model parameters are measured by autocorrelation, and MCMC will converge if the autocorrelation of each parameter gradually approaches zero during sampling \cite{sharma2017markov}.
By MCMC, the posterior distribution $p(\boldsymbol{\theta}|\boldsymbol{\mathrm{S}}_{N}^{M,\mathcal{K}^{(1)},\mathcal{K}^{(2)}})$ can be evaluated efficiently. 

\section{Numerical simulations}
\label{sec:simulation}
To verify the feasibility of the Bayesian spectral density approach \cite{yuen2010bayesian}, numerical simulations are carried out in this section.
\subsection{Predefined modal properties and FDs}
\begin{figure}[H]
	\vspace{-0.2cm}
    \centering
    \makebox[\textwidth][c]{\includegraphics[]{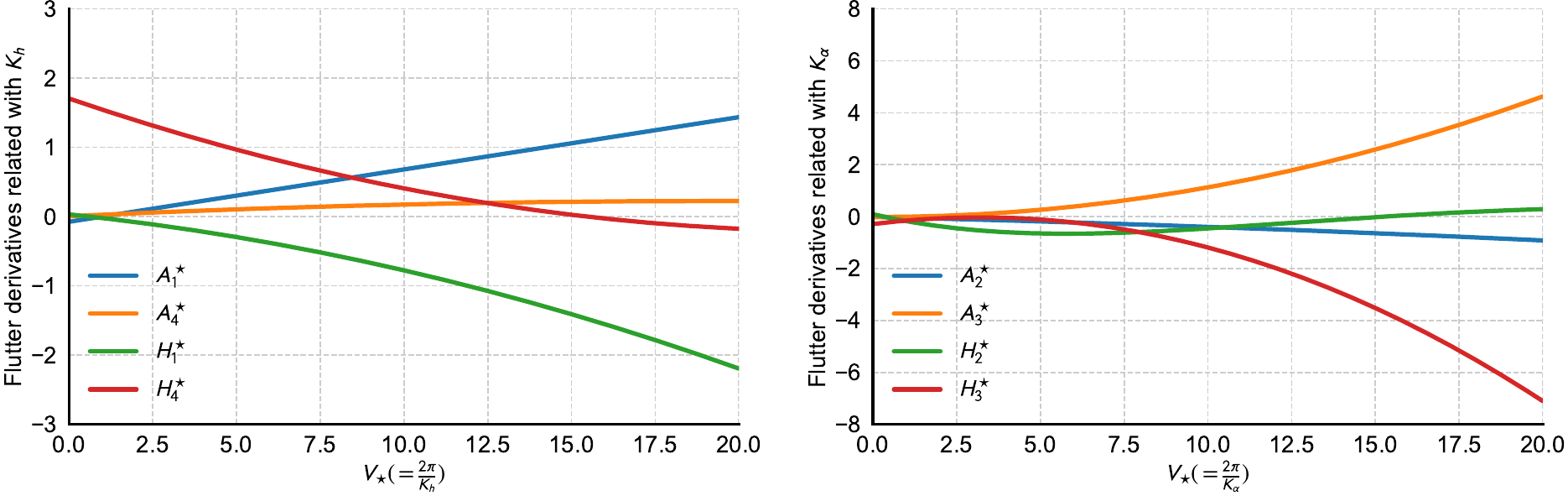}}
    \caption{Predefined flutter derivatives in the numerical simulations}
	\label{fig:FD_plot}
\end{figure}

The vertical and torsional frequencies are set as $f_{h}=0.1~\mathrm{Hz}$ and $f_{\alpha}=0.25~\mathrm{Hz}$, respectively. Both damping ratios are set as $0.005$. The bridge deck is $36~\mathrm{m}$. The mass per unit length is $27935~\mathrm{kg/m}$. The mass moment of inertia per unit length is $2595580~\mathrm{kg\times m^{2}/m}$. The air density is $1.225~\mathrm{kg/m^{3}}$. The predefined FDs are shown in Fig.~\ref{fig:FD_plot}, where $V_{\star}$ means the reduced velocity, $K_{h}$ and $K_{\alpha}$ are the reduced frequencies as described in Sec.~\ref{sec:spectral_density_formulation}. 

\subsection{Buffeting displacement response}
The buffeting forces $\boldsymbol{f}(t)$ in Eq.~\eqref{eq:lti_eq} are assumed as a white noise loading process, and the PSD matrix of the buffeting forces is 
\begin{equation}
\label{eq:wgn_buffeting}
\mathbf{S}_{\boldsymbol{f}}(\omega)=\mathbf{S}_{\boldsymbol{f}}=\left[\begin{array}{cc}
0.001 & 0 \\
0 & 0.001
\end{array}\right] N^{2} \cdot \mathrm{s}~\left(\omega > 0 \right)
\end{equation}

\begin{figure}[H]
    \centering
    \makebox[\textwidth][c]{\includegraphics[]{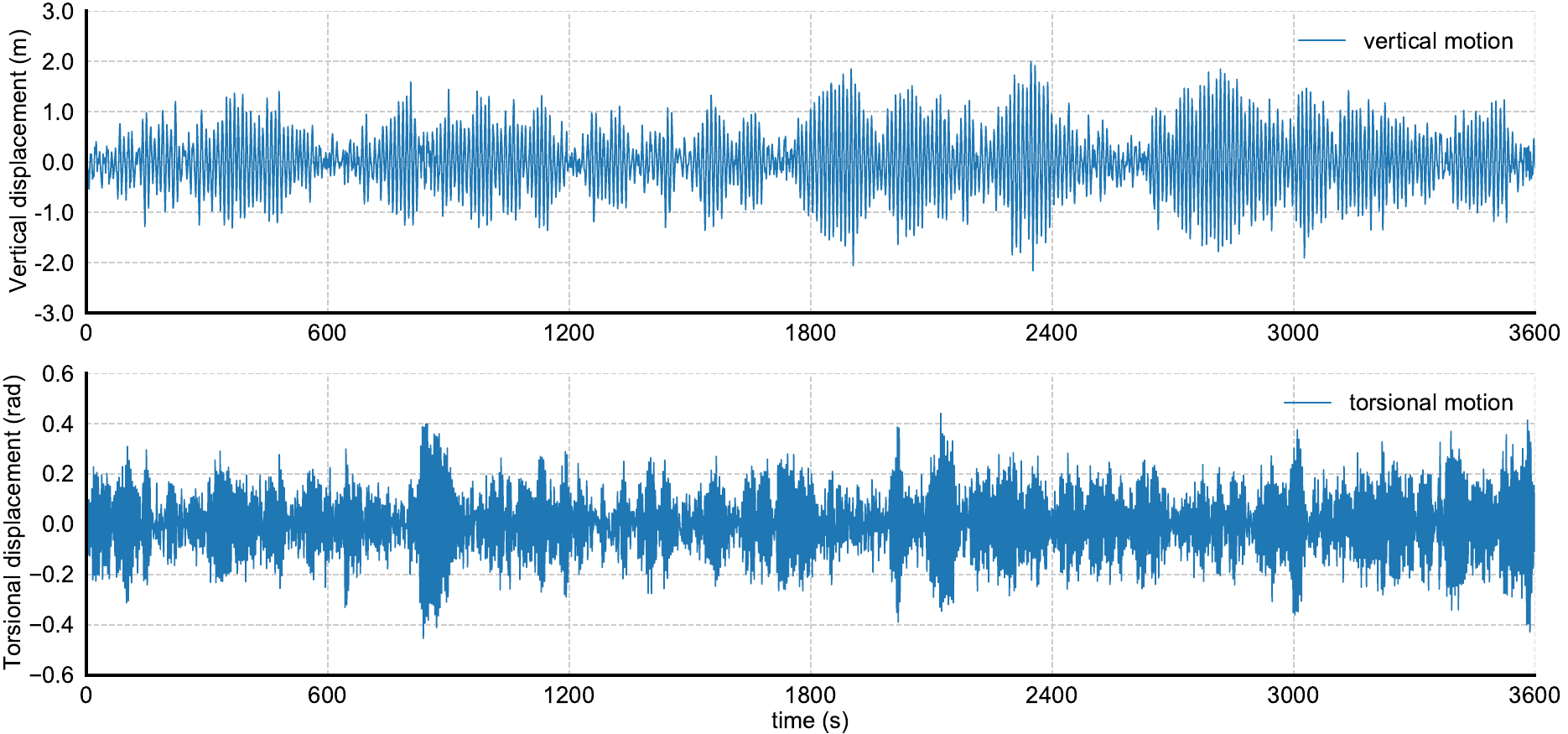}}
    \caption{Buffeting vertical displacement response and torsional displacement response when $U=30$~m/s ($K_{h} = 0.7540$, $K_{\alpha} = 1.8850$)}
	\label{fig:buffeting_response}
\end{figure}

The response time series are generated by ``lsim'' function with MATLAB, where the sampling interval is $0.01~\mathrm{s}$ (i.e., $f_{s}=100~\mathrm{Hz}$). Fig.~\ref{fig:buffeting_response} shows an example of vertical and torsional displacements when the mean wind speed $U=30~\mathrm{m/s}$ (i.e., $K_{h} = 0.7540$ and $K_{\alpha} = 1.8850$).

\begin{figure}[H]
    \centering
    \makebox[\textwidth][c]{\includegraphics[]{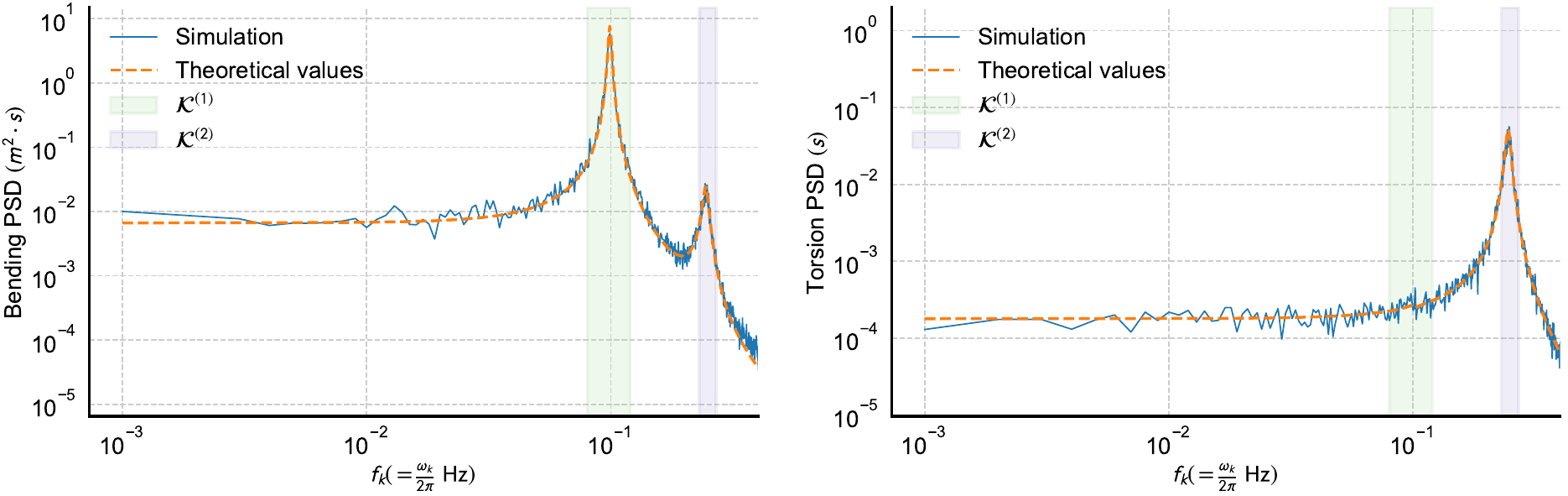}}
    \caption{Simulated and theoretical PSDs of buffeting displacement responses when $U=30$~m/s ($K_{h} = 0.7540$, $K_{\alpha} = 1.8850$)}
	\label{fig:buffeting_response_PSD}
\end{figure}

As shown in Fig.~\ref{fig:buffeting_response_PSD}, the selected frequency set $\mathcal{K}^{(1)}$ is $0.08\mathrm{Hz} \sim 0.12\mathrm{Hz}$, $\mathcal{K}^{(2)}$ is $0.23\mathrm{Hz} \sim 0.27\mathrm{Hz}$. According to Eq.~\eqref{eq:PSD_displacement}, we have the theoretical PSD of displacements. In Fig.~\ref{fig:buffeting_response_PSD}, the blue continuous line is plotted by the simulated signals with Eq.~\eqref{eq:PSD_sampling}, and the orange dashed line means the theoretically-calculated values with Eq.~\eqref{eq:PSD_displacement}. The $M$ in Eq.~\eqref{eq:PSD_sampling} is set as $18$. Fig.~\ref{fig:buffeting_response_PSD} shows that the PSDs of simulated buffeting responses are consistent with that of the theoretical values, which means that the simulated signals are correct. 

\subsection{MCMC sampling}
\subsubsection{Prior distributions}
The determination of the prior distributions is important for the calculation and convergence of posterior distributions \cite{yuen2010bayesian}. For FDs $A_{i}^{\star},H_{i}^{\star} (i=1,2,3,4)$ and modified buffeting force PSDs $S_{L,1}$,$S_{L,2}$,$S_{M,1}$,$S_{M,2}$, the prior distributions are assumed to obey the uniform distribution (i.e., there are no prior information) independently.


Bayesian inference is utilized to quantify uncertainties, and AIES-based MCMC sampling is adopted to estimate the posterior PDF of the input parameters \cite{goodman2010ensemble,grinsted2015gwmcmc}. AIES-based MCMC sampling is set to run $200,000$ samples, thinning by taking every 10th sample and a burn-in of $20\%$ total samples is employed.

\subsubsection{Convergence diagnostics}
\begin{figure}[H]
    \centering
    \makebox[\textwidth][c]{\includegraphics[]{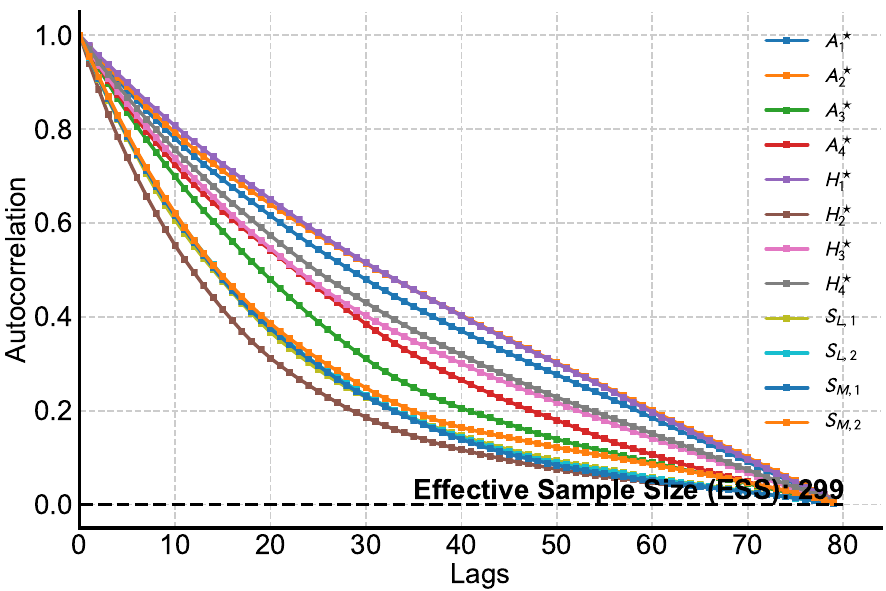}}
    \caption{MCMC autocorrelation (case of simulation)}
	\label{fig:MCMC_autocorrelation}
\end{figure}

To diagnose convergence of Markov chain, Fig.~\ref{fig:MCMC_autocorrelation} shows that the autocorrelations of the input parameters approach zero when $\mathrm{lags}=80$, which means that the Markov chains converge \cite{sharma2017markov}.

\subsubsection{Posterior distributions}
\begin{figure}[H]
    \centering
    \makebox[\textwidth][c]{\includegraphics[]{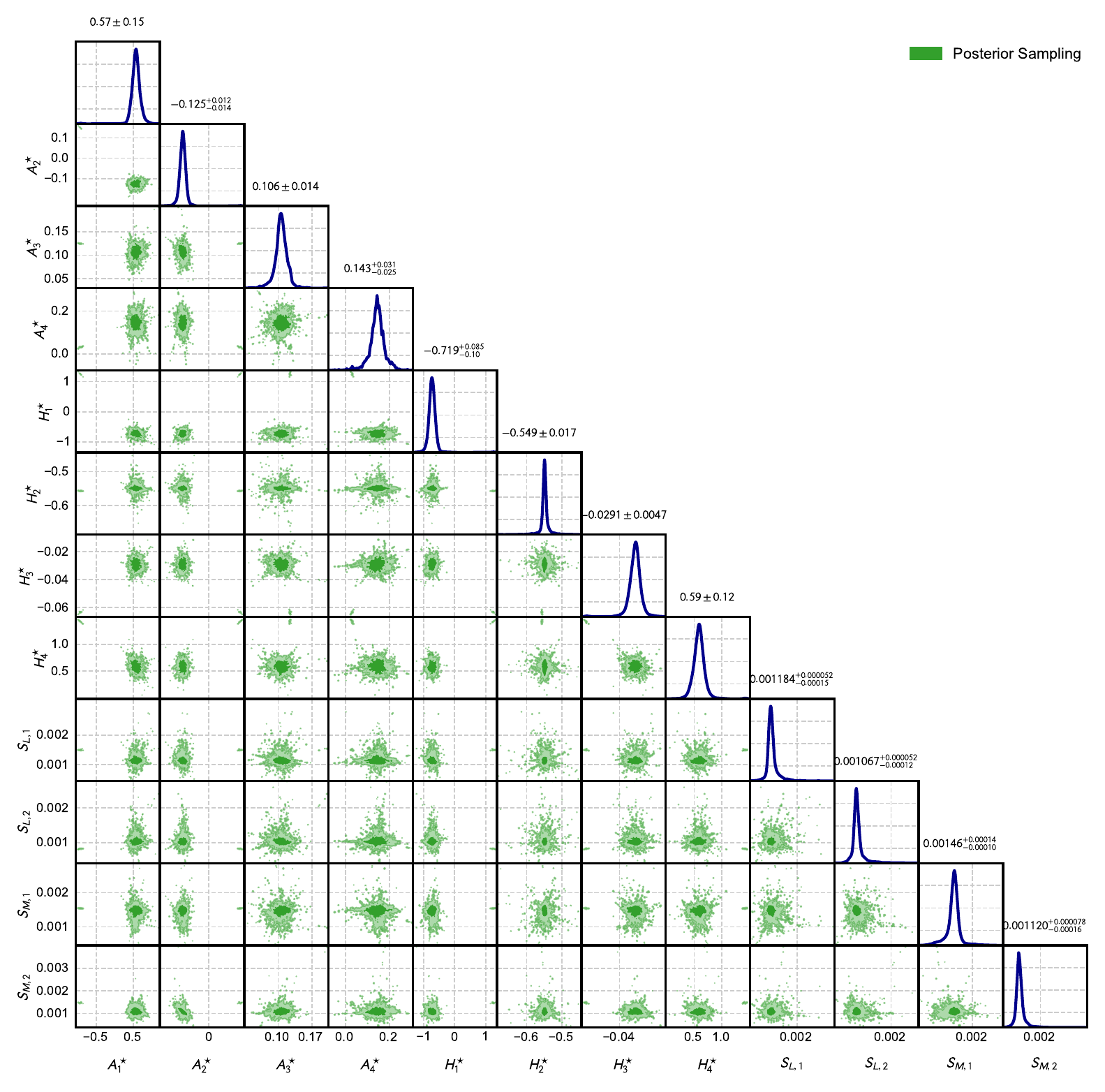}}
    \caption{Posterior distributions of FDs and buffeting force PSDs when $U=30$~m/s ($K_{h} = 0.7540$, $K_{\alpha} = 1.8850$)}
	\label{fig:MCMC_sampling_simu}
\end{figure}

Fig.~\ref{fig:MCMC_sampling_simu} illustrates the posterior distributions of $\mathbf{\Theta}$. The numerical simulation shows that if the cross spectral of buffeting PSDs is zero as the preassumption in Eq.~\eqref{eq:wgn_buffeting} (i.e., lift buffeting force is independent upon moment buffeting force), FDs ($A_{i}^{\star},H_{i}^{\star},i=1,2,3,4$) are approximately mutually independent. Since this study focuses on FDs, the following content will not discuss the MPVs and corresponding uncertainty of buffeting force PSDs.

\subsubsection{Marginal distributions of flutter derivatives and reconstruction of buffeting responses' PSDs}
\begin{figure}[H]
    \centering
    \makebox[\textwidth][c]{\includegraphics[]{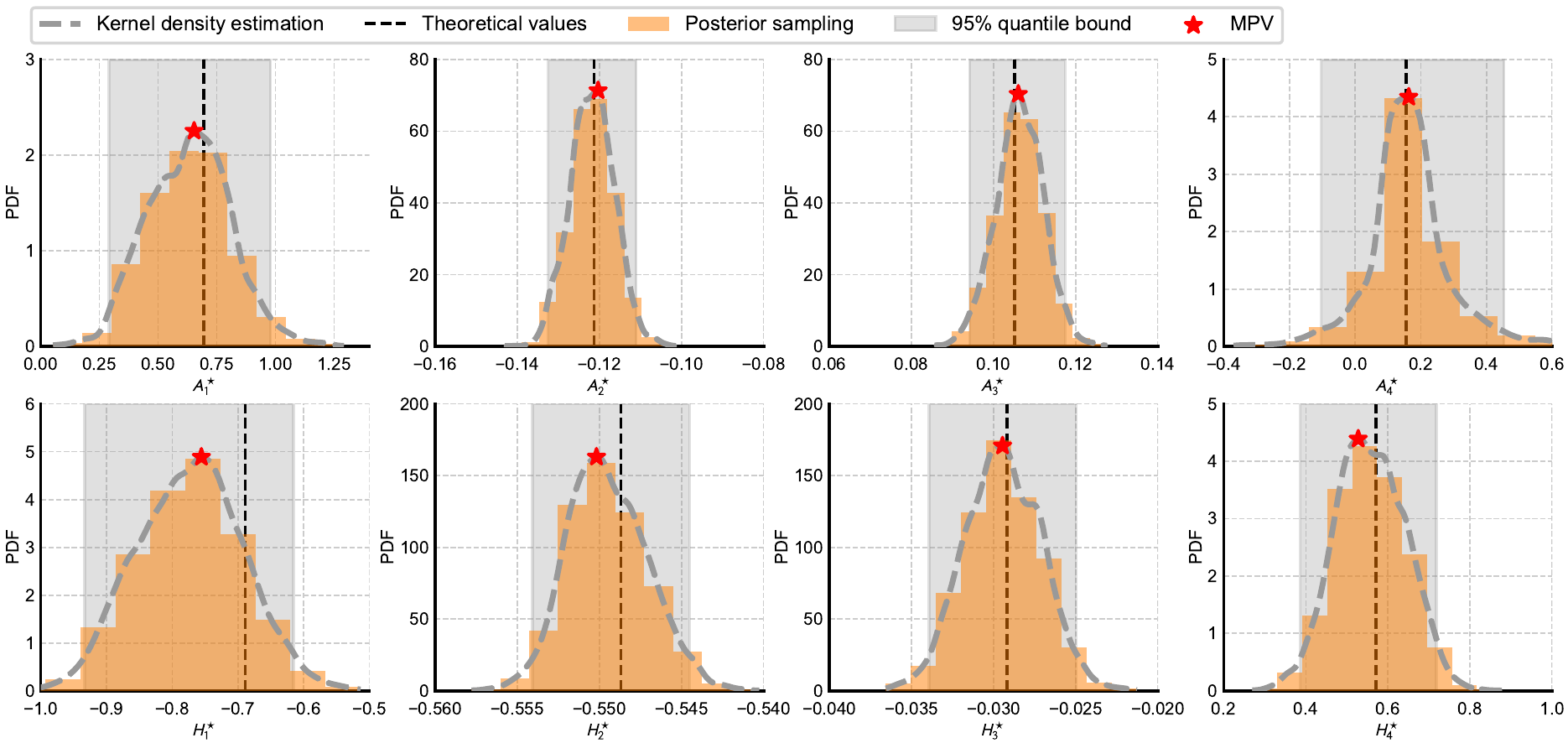}}                                         
    \caption{Posterior marginal distributions of FDs when $U=30$~m/s ($K_{h} = 0.7540$, $K_{\alpha} = 1.8850$)}
	\label{fig:prior_posterior_1d}
\end{figure}

Fig.~\ref{fig:prior_posterior_1d} shows the posterior marginal distributions and $95\%$ quantile bounds, where the kernel density estimation (KDE) is adopted to get the MPVs (position with the largest probability) \cite{sheather1991reliable}. The MPVs are consistent with the theoretical values generally for each input parameter, where the small deviations that exist in $H_{1}^{\star}$ and $H_{2}^{\star}$ may be caused by aliasing and leakage of the simulated signals \cite{yuen2010bayesian,au2017operational}. 

\begin{figure}[H]
    \centering
    \makebox[\textwidth][c]{\includegraphics[]{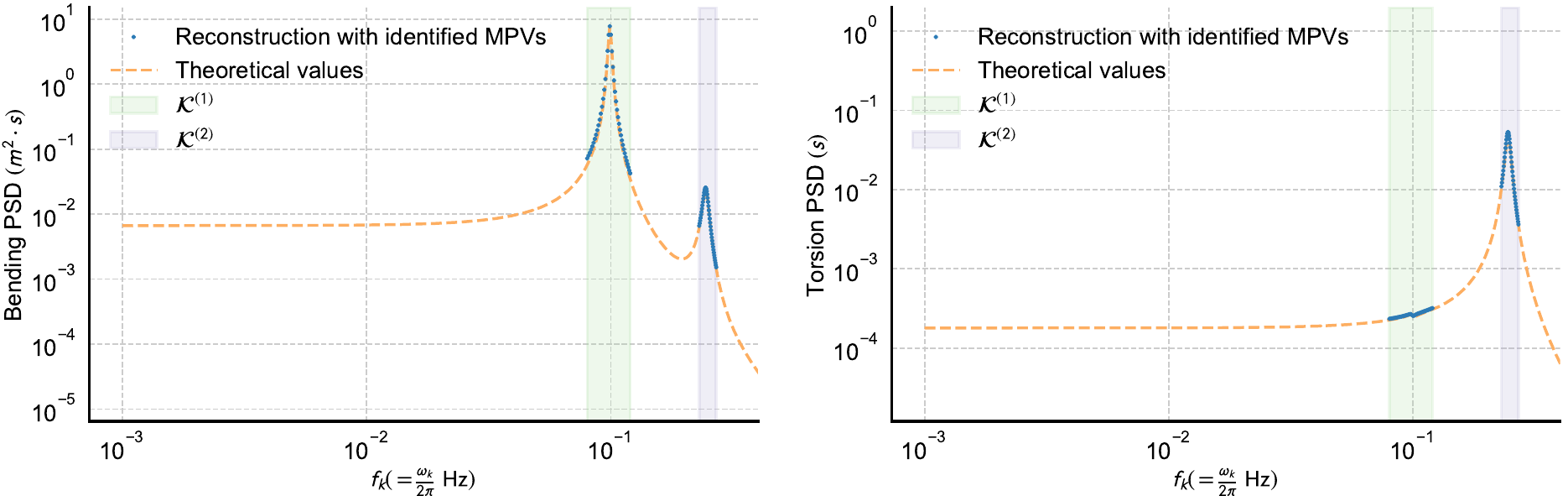}}
    \caption{Comparison between the reconstructed PSDs (with identified MPVs of $A_{i}^{\star}$, $H_{i}^{\star}$, $i=1,2,3,4$, and $S_{L,1}$, $S_{L,2}$, $S_{M,1}$, $S_{M,2}$) and the theoretical PSDs when $U=30$~m/s ($K_{h} = 0.7540$, $K_{\alpha} = 1.8850$)}
	\label{fig:theo_identified_PSD}
\end{figure}

Fig.~\ref{fig:theo_identified_PSD} shows the reconstruction of buffeting responses' PSDs with the identified MPVs of $A_{i}^{\star}$, $H_{i}^{\star}$ ($i=1,\dots,4$), $\mathrm{S}_{L,1}$, $S_{L,2}$, $S_{M,1}$ and $\mathrm{S}_{M,2}$. It is clear that though some small deviation exists in Fig.~\ref{fig:prior_posterior_1d}, the reconstructed PSDs are still consistent with the theoretical ones.

\subsection{Most probable values (MPVs) of FDs at various reduced velocities}
\begin{figure}[!htb]
    \centering
    \makebox[\textwidth][c]{\includegraphics[]{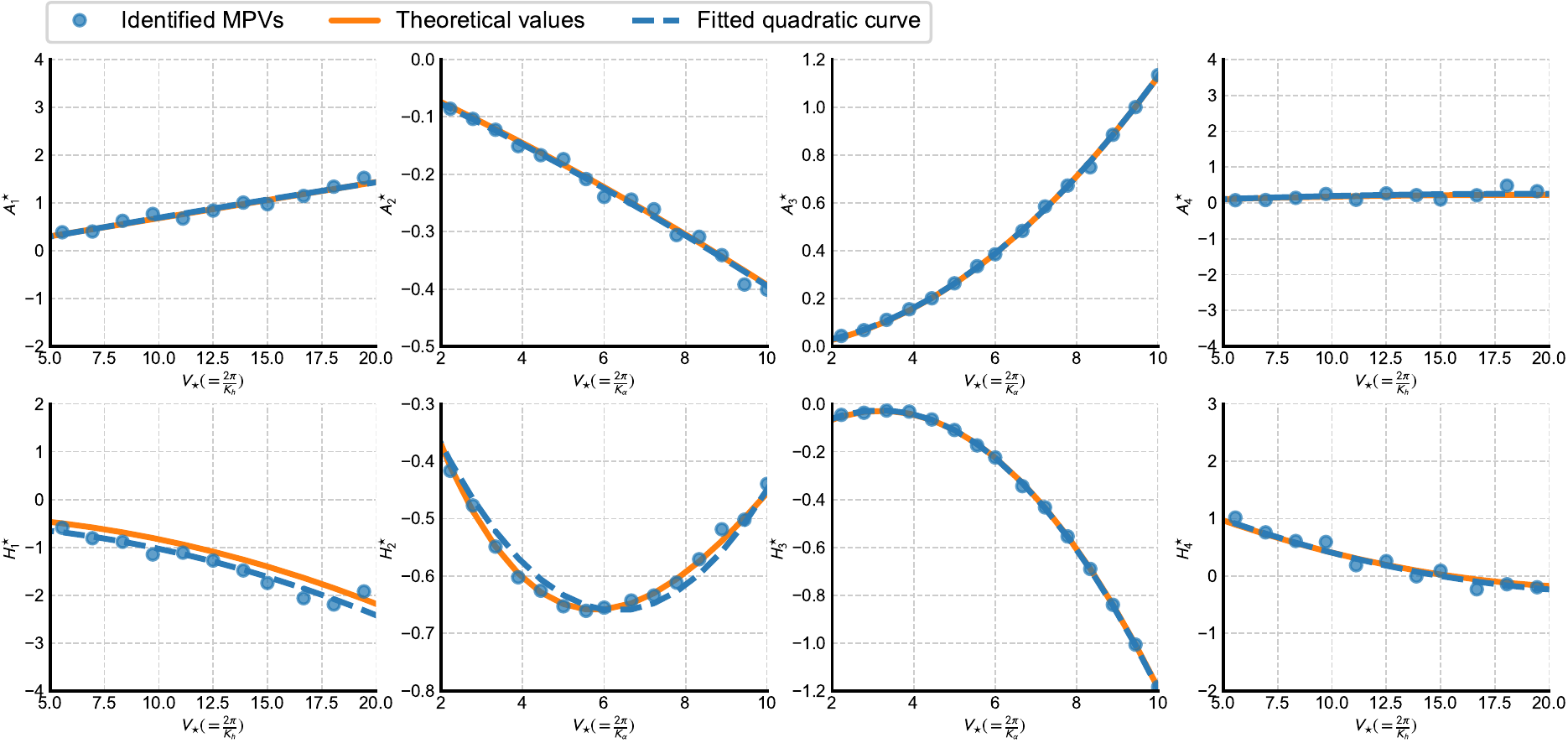}}
    \caption{MPVs and corresponding uncertainty of identified FDs at various reduced velocities}
	\label{fig:simu_MPV_uq}
\end{figure}

Fig.~\ref{fig:simu_MPV_uq} shows the comparison between identified FDs' MPVs and corresponding predefined theoretical values. The identified MPVs are consistent with the predefined theoretical values at various reduced velocities in general. 

\section{Wind tunnel tests}
To evaluate the feasibility of Bayesian spectral density approach in extracting FDs of bridge sectional model in turbulent flow, the wind tunnel tests of a streamlined thin plate model and a center-slotted girder sectional model are carried out, respectively. The thin plate is employed because there is theoretical solutions, i.e., Theodorsen functions \cite{scanlan1993problematics}, of this shape. The center-slotted girder section is employed because this kind of cross sectional shape is widely utilized in current engineering practice for the construction of long-span bridges. The wind tunnel tests are carried out in TJ-1 wind tunnel at Tongji University in China. TJ-1 is an open straight-through-type boundary layer wind tunnel with a working section of $1.8$~m(width)$\times1.8$~m(height), and the wind speed can be continuously operated within the range of $0.5\sim 25~\mathrm{m/s}$.

\subsection{Case 1: thin plate model}
The FDs of a thin plate can be derived theoretically (i.e., Theodorsen function \cite{scanlan1993problematics}). As a result, a thin plate model is tested to verify the Bayesian spectral density approach, where the Theodorsen results are given as follows \cite{scanlan1993problematics}:
\begin{subequations}
\small
\label{eq:FD_plate}
\begin{equation}
H_{1}^{*}=\frac{-\pi F(k)}{2 k}, H_{2}^{*}=\frac{-\pi}{8 k}\left[1+F(k)+\frac{2 G(k)}{k}\right], H_{3}^{*}=\frac{-\pi}{4 k^{2}}\left[F(k)-\frac{k \cdot G(k)}{2}\right], H_{4}^{*}=\frac{\pi}{4}\left[1+\frac{2 G(k)}{k}\right]
\end{equation}
\begin{equation}
A_{1}^{*}=\frac{\pi F(k)}{8 k}, A_{2}^{*}=\frac{-\pi}{32 k}\left[1-F(k)-\frac{2 G(k)}{k}\right], A_{3}^{*}=\frac{\pi}{16 k^{2}}\left[\frac{k^{2}}{8}+F(k)-\frac{k \cdot G(k)}{2}\right], A_{4}^{*}=\frac{-\pi G(k)}{8 k}
\end{equation}
\begin{equation}
F(k)=1-\frac{0.165}{1+\left(\frac{0.0455}{k}\right)^{2}}-\frac{0.335}{1+\left(\frac{0.3}{k}\right)^{2}},~ G(k)=-\frac{\frac{0.165 \times 0.0455}{k}}{1+\left(\frac{0.0455}{k}\right)^{2}}-\frac{\frac{0.335 \times 0.3}{k}}{1+\left(\frac{0.3}{k}\right)^{2}}
\end{equation}
\end{subequations}
where $k=\frac{\omega B}{2 U}=\frac{K}{2}$, $B$ is the width of the sectional model, $K$ is the reduced frequency.

\begin{figure}[H]
    \centering
    \makebox[\textwidth][c]{
    \begin{subfigure}[c]{90mm}
      \centering
      \includegraphics[scale=0.13]{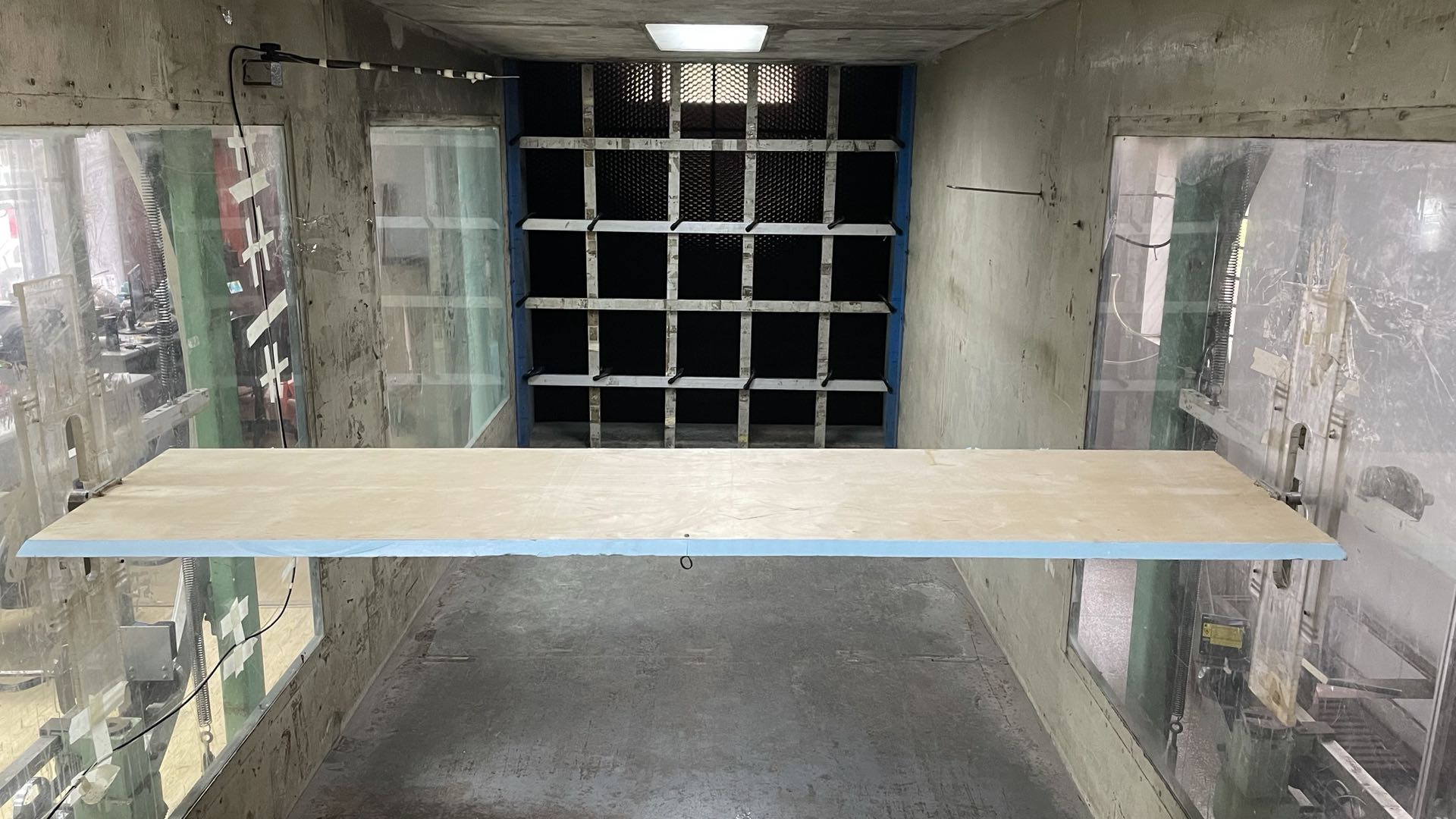}
      \caption{Thin plate model with regularly-arranged grids}
      \label{fig:grid_wind_tunnel}
    \end{subfigure}
    \begin{subfigure}[c]{90mm}
      \centering
      \includegraphics[scale=0.48]{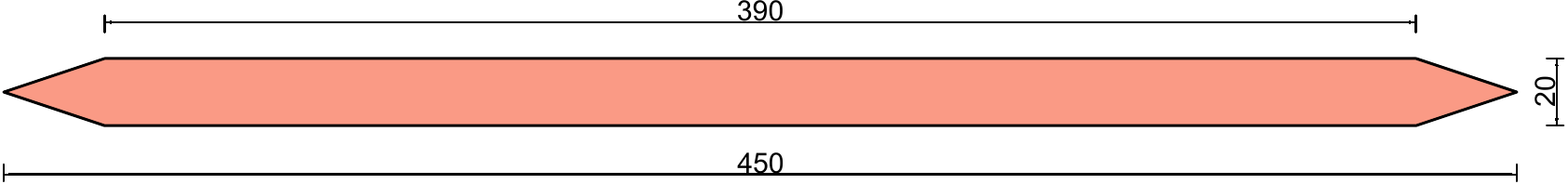}
      \caption{Cross sectional shape of the thin plate model}
	  \label{fig:thin_plate_model}
    \end{subfigure}
    }
    \caption{Case 1: the thin plate model}
\end{figure}

As shown in Fig.~\ref{fig:grid_wind_tunnel}, the turbulent flow was generated by regularly-arranged grids, where the thin plate model was suspended by eight springs. The thin plate model is centrally-symmetric with $1.745~\mathrm{m}$ in the longitudinal dimension. As shown in Fig.~\ref{fig:thin_plate_model}, the size of the cross section is $450~\mathrm{mm}\times 20~\mathrm{mm}$ (length$\times$height). The mass is $6~\mathrm{kg/m}$; the mass moment of inertia is $0.7\mathrm{kg\times m^{2}/m}$; the vertical modal frequency is $1.9~\mathrm{Hz}$, the vertical damping ratio is $0.004$; the torsional modal frequency is $3.05~\mathrm{Hz}$, the torsional damping ratio is $0.003$.

The detailed identification process has been discussed in Sec.~\ref{sec:simulation}. In this subsection, we take the condition of the mean wind speed $U=8.6~\mathrm{m/s}$ for example to explain the identification process of the thin plate. The sampling rate is $1024~\mathrm{Hz}$. Conventionally with the previous least square method \cite{gu2000identification} in the free vibration test, the sampling duration is $10~\mathrm{s} \sim 20~\mathrm{s}$. When utilizing the buffeting response, the sampling duration is usually much longer (e.g., $14~\mathrm{min}$ with stochastic subspace identification technology in \cite{qin2004determination}). In order to obtain reliable and stable measured PSDs, the sampling duration in the case of thin plate model is $2000~\mathrm{s}$, $M$ in Eq.~\eqref{eq:complex_Wishart_distribution} is $8$.

\begin{figure}[H]
    \centering
    \makebox[\textwidth][c]{\includegraphics[]{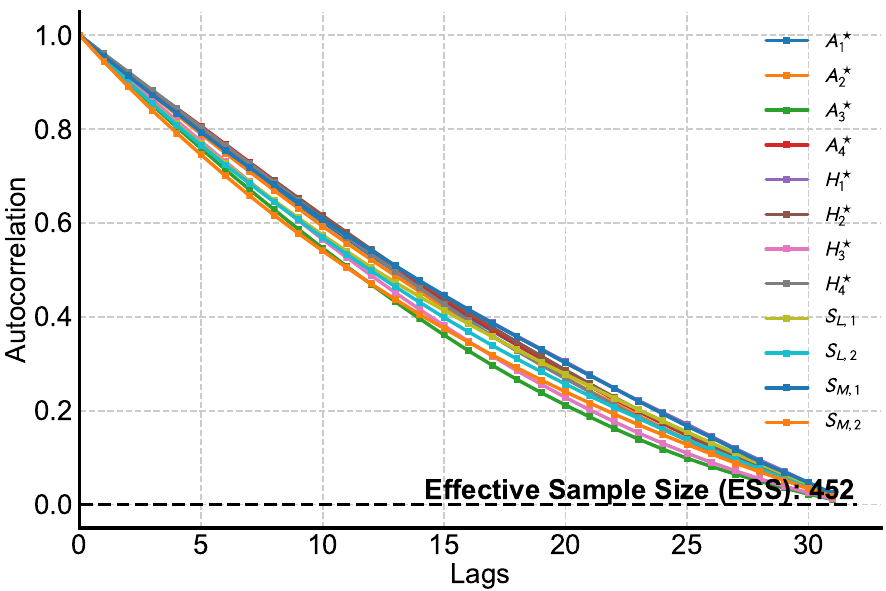}}
    \caption{Case 1: MCMC autocorrelation}
	\label{fig:MCMC_autocorrelation_plate}
\end{figure}

Fig.~\ref{fig:MCMC_autocorrelation_plate} shows that the Markov chain has converged since the autocorrelation of each parameter in $\boldsymbol{\theta}$ approaches zero when $\mathrm{lags}=31$.

\begin{figure}[H]
    \centering
    \makebox[\textwidth][c]{\includegraphics[]{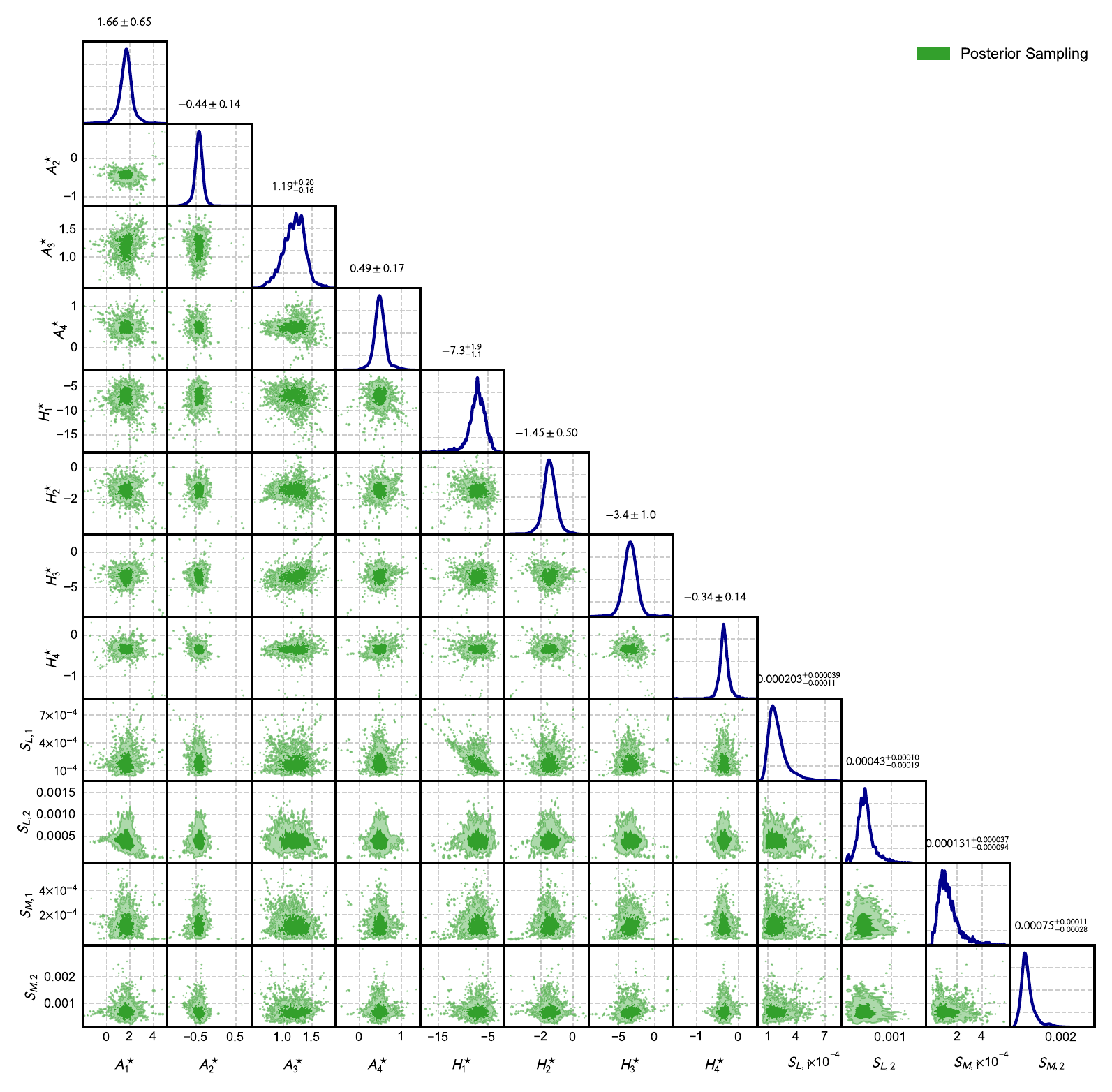}}
    \caption{Case 1: posterior distributions of flutter derivatives of the thin plate when $U=8.6$~m/s ($K_{h}=0.6247$, $K_{\alpha}=1.0028$)}
	\label{fig:MCMC_sampling_simu_plate}
\end{figure}

As shown in Fig.~\ref{fig:MCMC_sampling_simu_plate}, for the thin plate, the FDs are approximately independent with each other in a single test. It should be stressed again that the variances shown in each posterior FD sampling, namely identification uncertainty here, are different from the experimental uncertainties examined in \cite{seo2011estimation,fang2020experimental}. The variances of each FD in this study actually stand for the posterior marginal distributions, which heavily depends on the prior information \cite{yuen2010bayesian}. We can also obtain the experimental uncertainties through the proposed Bayesian spectral density approach by statistical analysis of the FDs' MPVs in many tests with the same preset experimental conditions. Furthermore, the correlation of each FD in this study is also different from that of experimental uncertainties due to the same reason.  

\begin{figure}[H]
    \centering
    \makebox[\textwidth][c]{\includegraphics[]{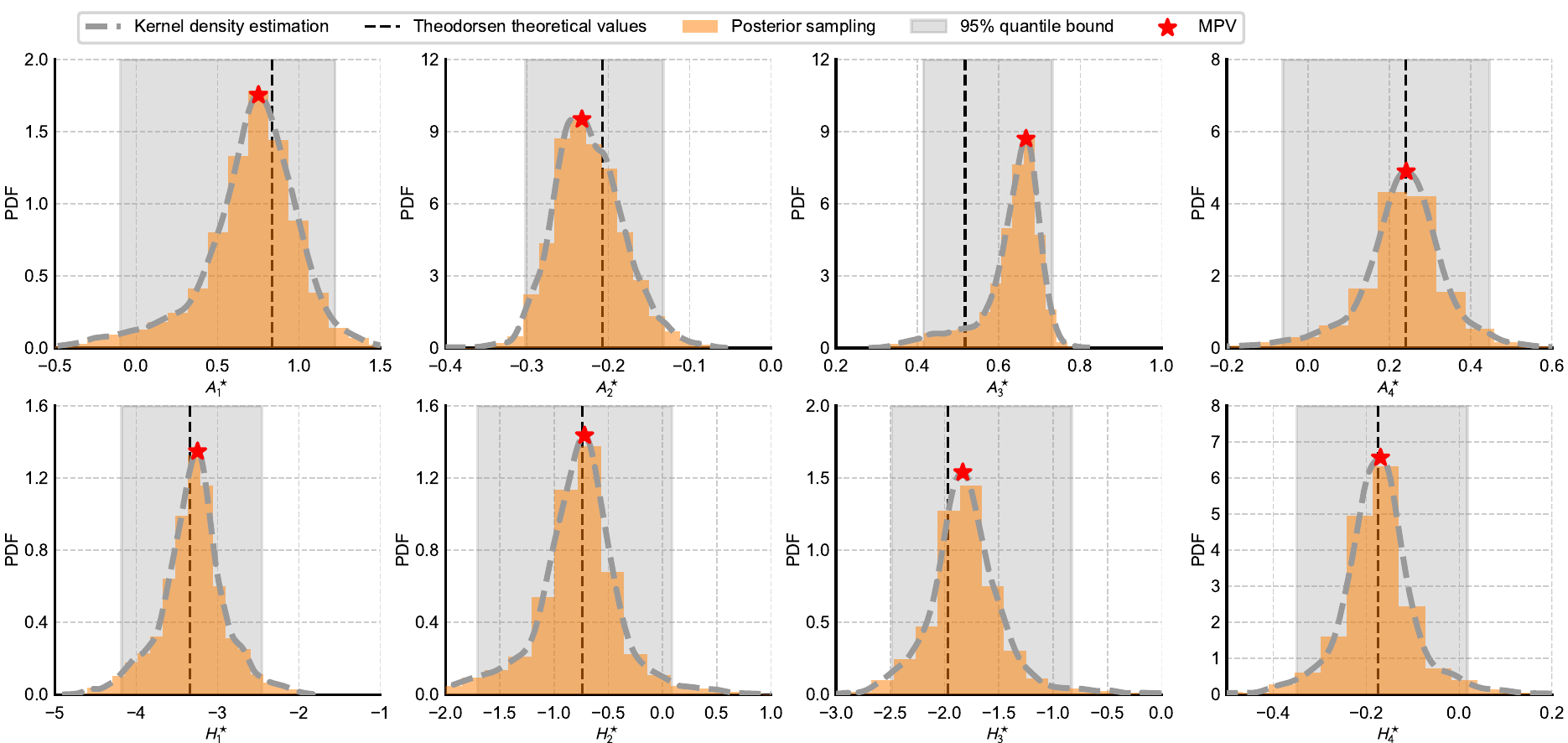}}                                         
    \caption{Case 1: posterior marginal distributions of flutter derivatives of the thin plate when $U=8.6$~m/s ($K_{h}=0.6247$, $K_{\alpha}=1.0028$)}
	\label{fig:prior_posterior_1d_plate}
\end{figure}

Fig.~\ref{fig:prior_posterior_1d_plate} shows the posterior marginal distributions of each FD, where the $95\%$ quantile bounds are also given. The dashed vertical line means the Theodorsen theoretical values, we can notice that the identified MPVs are generally consistent with the Theodorsen values. Some bias is observed in $A_{3}^{\star}$, reason of which is perhaps that the thin plate is not the rigorous infinite thin plate so the ``true'' value is not strictly the Theodorsen solution.

\begin{figure}[H]
    \centering
    \makebox[\textwidth][c]{\includegraphics[]{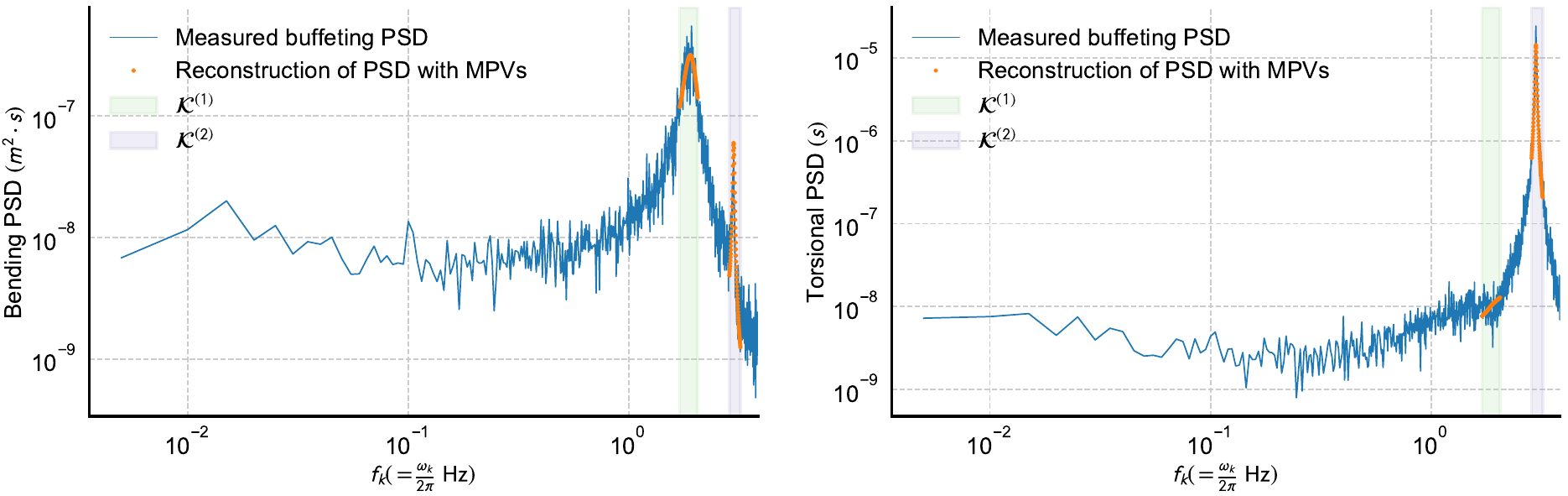}}
    \caption{Case 1: measured buffeting displacement PSDs and reconstruction of PSDs with identified MPVs of FDs and buffeting force PSDs when $U=8.6$~m/s ($K_{h}=0.6247$, $K_{\alpha}=1.0028$)}
	\label{fig:plate_PSD_reconstruction}
\end{figure}

Fig.~\ref{fig:plate_PSD_reconstruction} shows the reconstructed buffeting displacement PSDs with identified MPVs of FDs and buffeting force PSDs. The reconstructed PSDs are consistent with the measured ones, which can also prove that the identified FDs are correct. Specially, for a general bridge cross section (i.e., there are no theoretical values of FDs for reference), the most effective method to validate the identified MPVs is to reconstruct the buffeting displacement PSDs and compare those with the measured ones. 

\begin{figure}[!htb]
    \centering
    \makebox[\textwidth][c]{\includegraphics[]{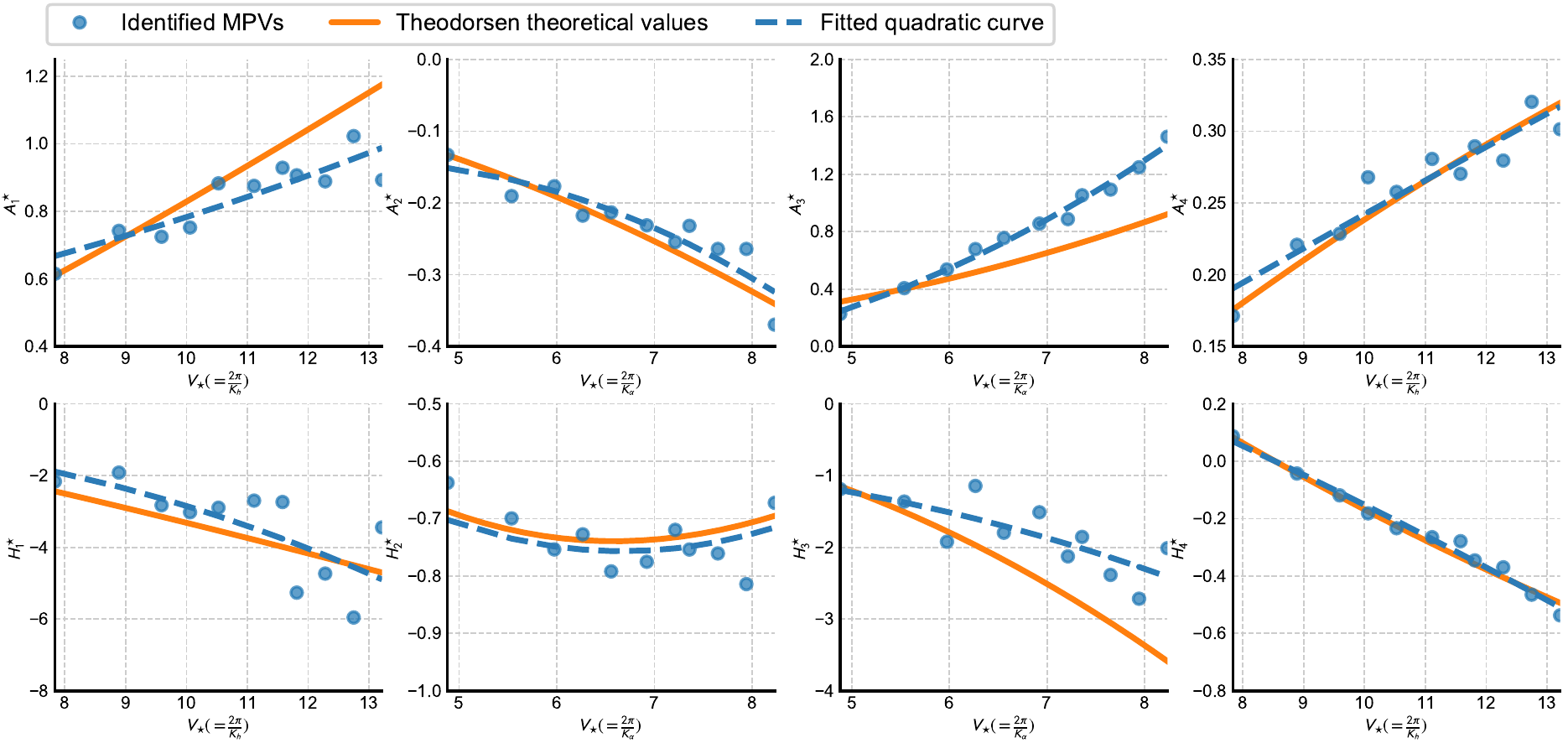}}
    \caption{Case 1: identified MPVs and corresponding uncertainty of flutter derivatives of the thin plate at various reduced velocities}
	\label{fig:simu_MPV_uq_plate}
\end{figure}

Fig.~\ref{fig:simu_MPV_uq_plate} shows the identified MPVs of FDs at various reduced velocities. The blue dashed line is the fitted quadratic curve with identified MPVs of FDs. Generally, the identified MPVs of FDs are consistent with the Theodorsen theoretical values. Again, identified $A_{3}^{\star}$ and $H_{3}^{\star}$ are relatively higher than the Theodorsen values since the thin plate model is not perfectly infinite. Similar bias can also be found when the FDs are extracted by stochastic subspace identification technology \cite{gu2004direct}.

\subsection{Case 2: sectional model of a center-slotted girder}
\begin{figure}[H]
    \centering
    \makebox[\textwidth][c]{
    \begin{subfigure}[c]{90mm}
      \centering
      \includegraphics[scale=0.06]{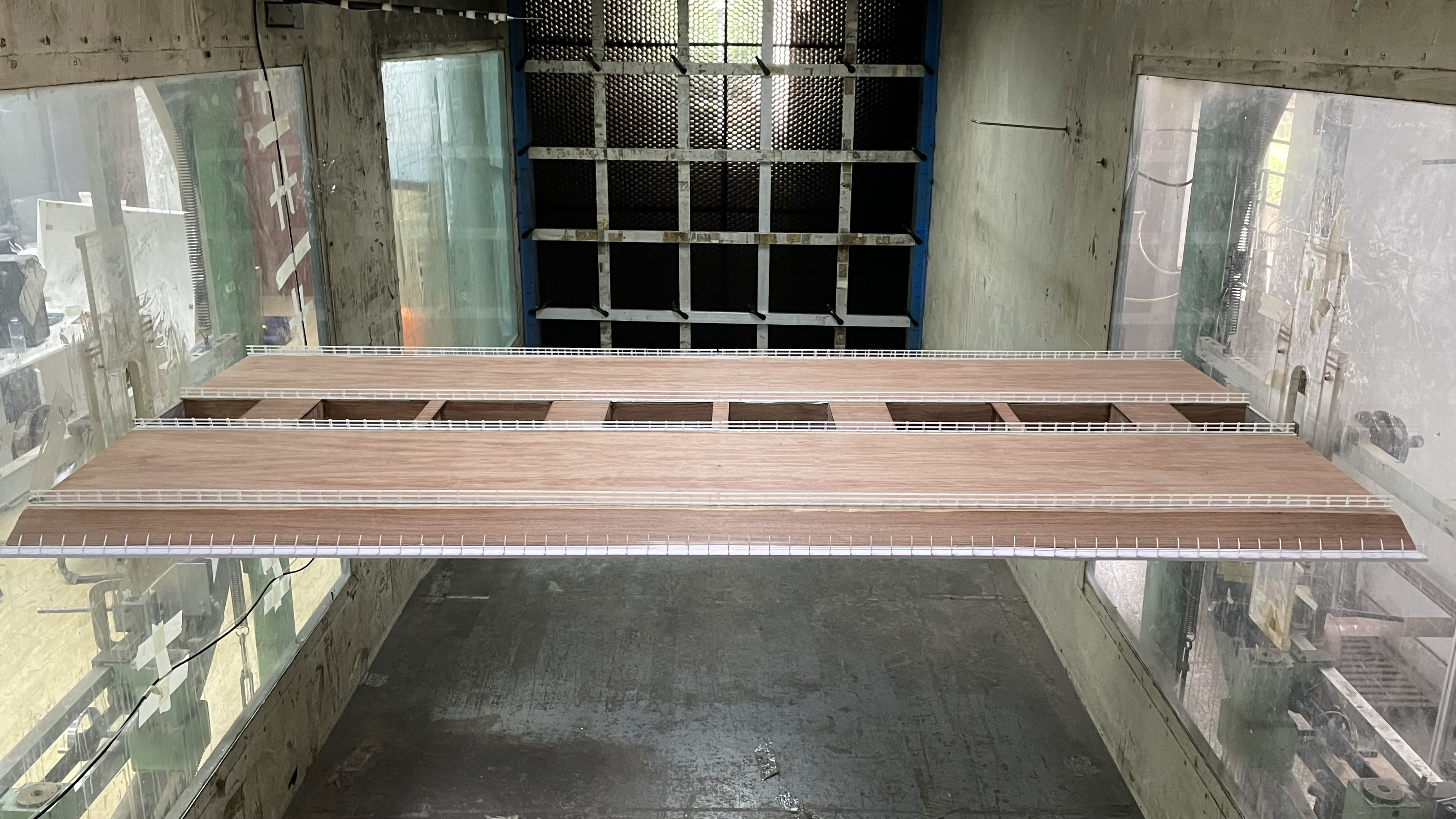}
      \caption{Center-slotted girder model with regularly-arranged grids}
      \label{fig:grid_XHM}
    \end{subfigure}
    \begin{subfigure}[c]{90mm}
      \centering
      \includegraphics[scale=0.6]{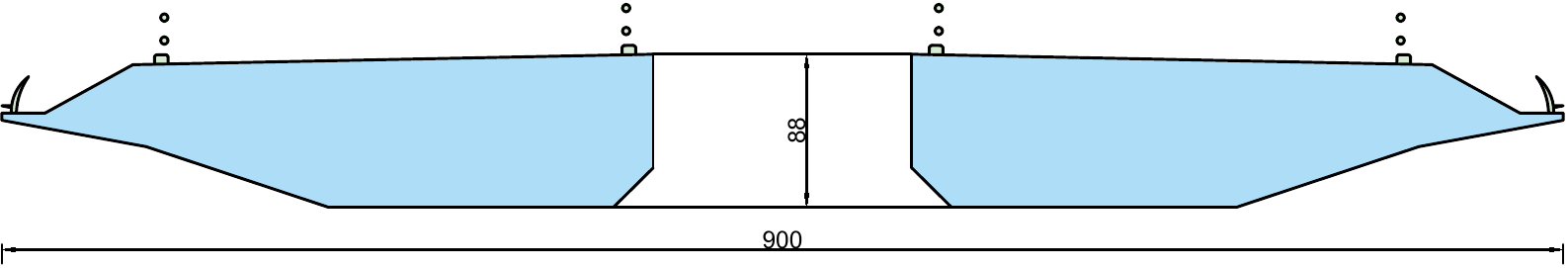}
      \caption{Cross sectional shape of the center-slotted girder model}
      \label{fig:XHM_cad}
    \end{subfigure}
    }
    \caption{Case 2: the center-slotted girder model}
    \label{fig:XHM_section}
\end{figure}

In order to investigate the applicability of the proposed Bayesian spectral density approach towards the general bridge cross section, we choose the widely-utilized center-slotted girder. Fig.~\ref{fig:XHM_section} shows the cross section of the center-slotted girder. It is $1.745~\mathrm{m}$ in the longitudinal dimension. The size of the cross section is $900~\mathrm{mm} \times 88~\mathrm{mm}$ (length$\times$height). The mass is $11.7~\mathrm{kg/m}$; the mass moment of inertia is $1.7~\mathrm{kg\times m^{2}/m}$; the vertical modal frequency is $1.86~\mathrm{Hz}$, the vertical damping ratio is $0.002$; the torsional modal frequency is $3.00~\mathrm{Hz}$, the torsional damping ratio is $0.0015$. The sampling rate is $1024~\mathrm{Hz}$. The sampling duration in the case of center-slotted girder model is $2000~\mathrm{s}$, $M$ in Eq.~\eqref{eq:complex_Wishart_distribution} is $8$.

\begin{figure}[H]
    \centering
    \makebox[\textwidth][c]{\includegraphics[]{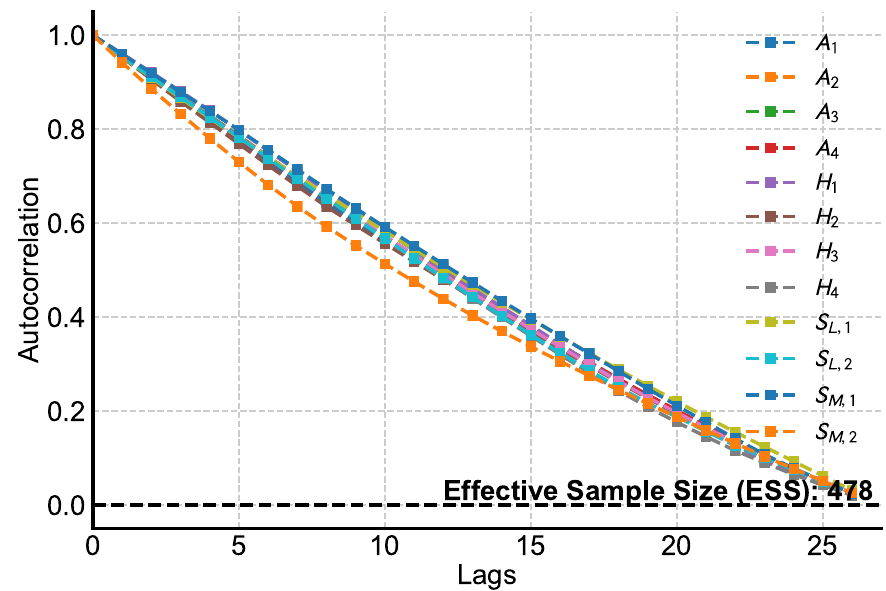}}
    \caption{Case 2: MCMC autocorrelation}
	\label{fig:MCMC_autocorrelation_XHM}
\end{figure}

In the case of center-slotted girder model, take the mean wind speed $U=7.25~\mathrm{m/s}$ for example. Fig.~\ref{fig:MCMC_autocorrelation_XHM} shows that the autocorrelation approaches zero when $\mathrm{lags}=27$, which means the Markov chain converges.

\begin{figure}[H]
    \centering
    \makebox[\textwidth][c]{\includegraphics[scale=0.98]{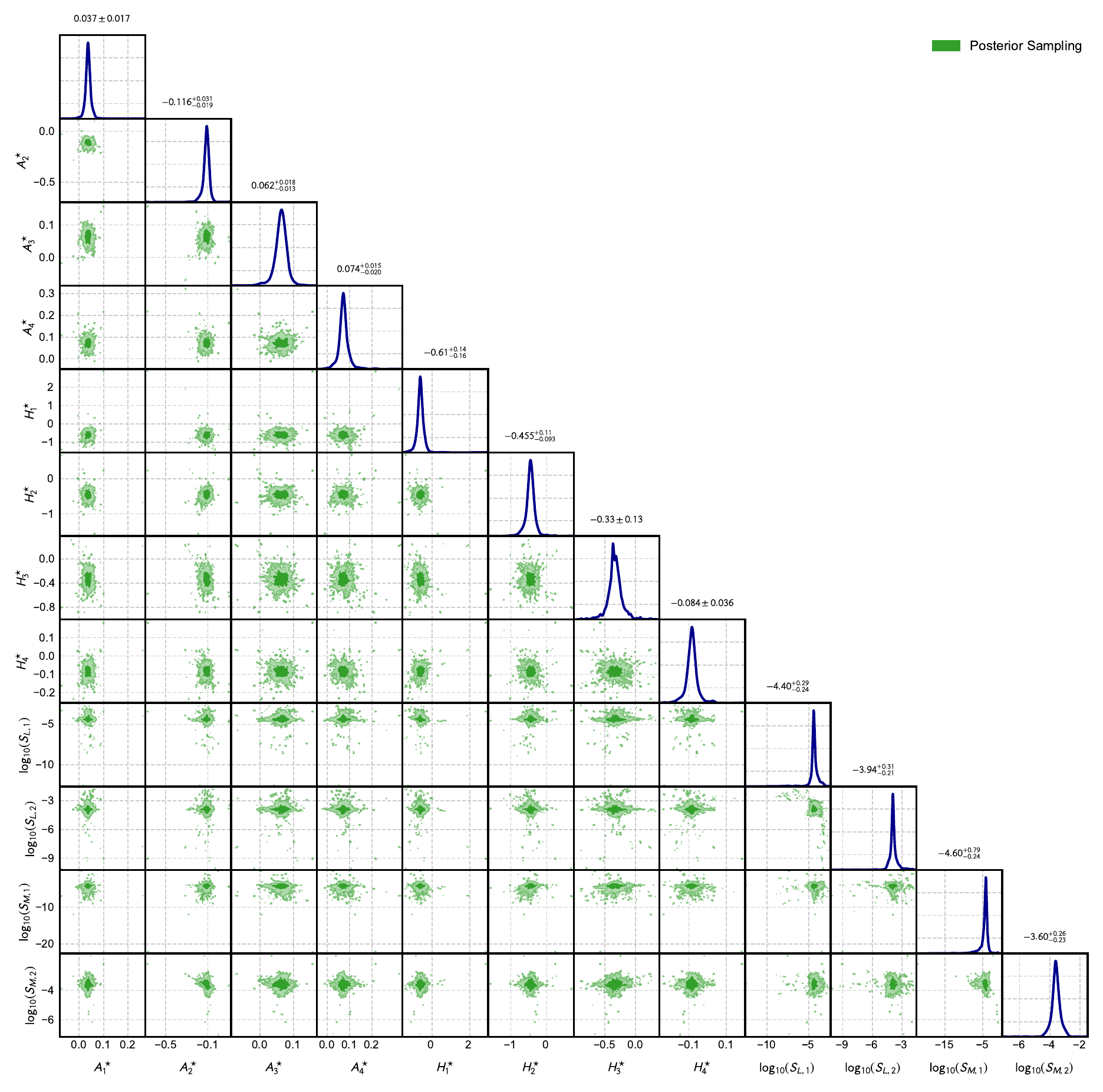}}
    \caption{Case 2: posterior distributions of flutter derivatives and buffeting force PSDs of the center-slotted girder model when $U=7.25~\mathrm{m/s}$ ($K_{h}=1.4547,K_{\alpha}=2.3399$)}
	\label{fig:MCMC_sampling_simu_XHM}
\end{figure}

Fig.~\ref{fig:MCMC_sampling_simu_XHM} shows the posterior samplings, where the FDs are independent with each other in a single test.

\begin{figure}[H]
    \centering
    \makebox[\textwidth][c]{\includegraphics[]{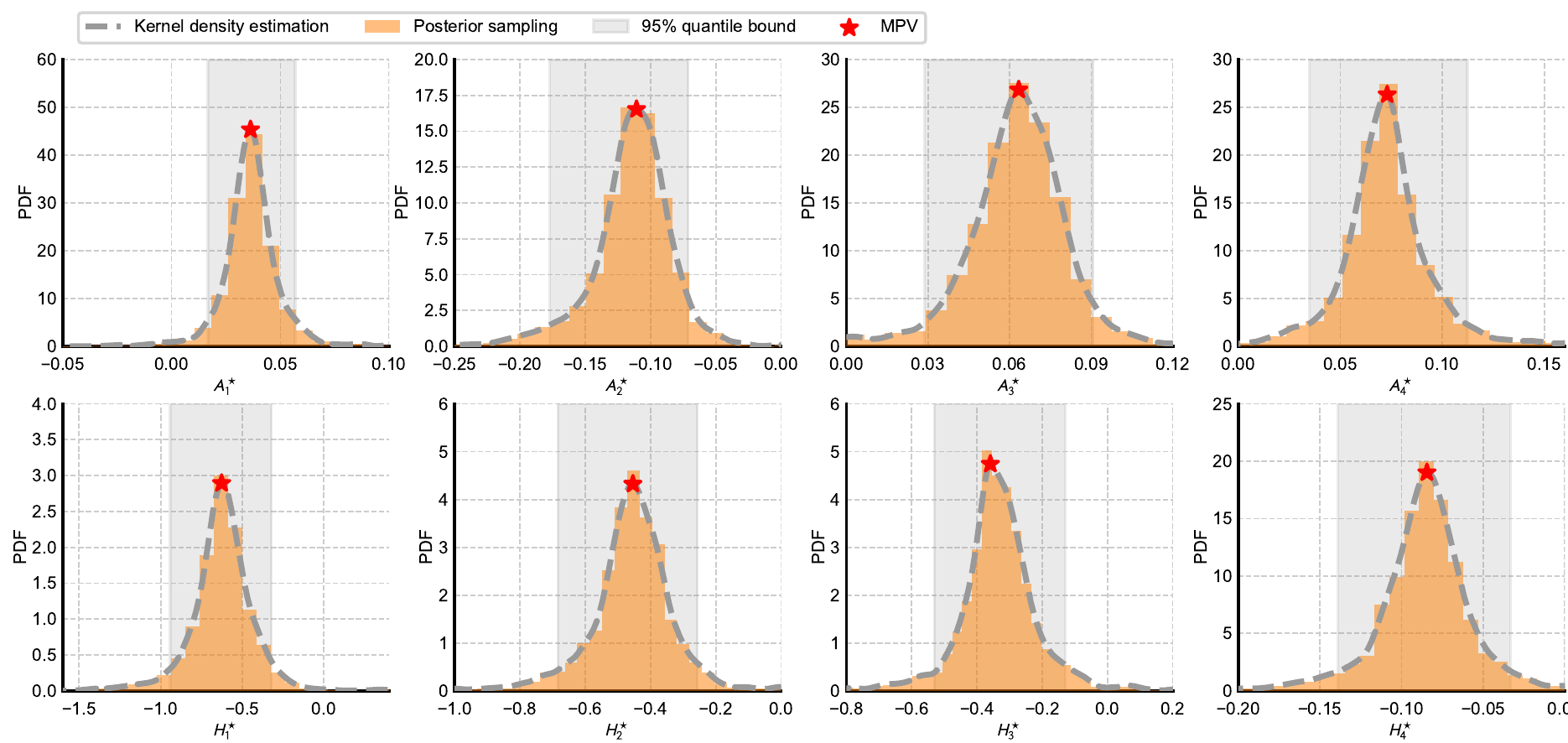}}                                         
    \caption{Case 2: posterior marginal distribution of flutter derivatives of the center-slotted girder model when $U=7.25~\mathrm{m/s}$ ($K_{h}=1.4547,K_{\alpha}=2.3399$)}
	\label{fig:prior_posterior_1d_XHM}
\end{figure}

Fig.~\ref{fig:prior_posterior_1d_XHM} shows the posterior marginal distributions of FDs, where the $95\%$ quantile bounds are also given.

\begin{figure}[H]
    \centering
    \makebox[\textwidth][c]{\includegraphics[]{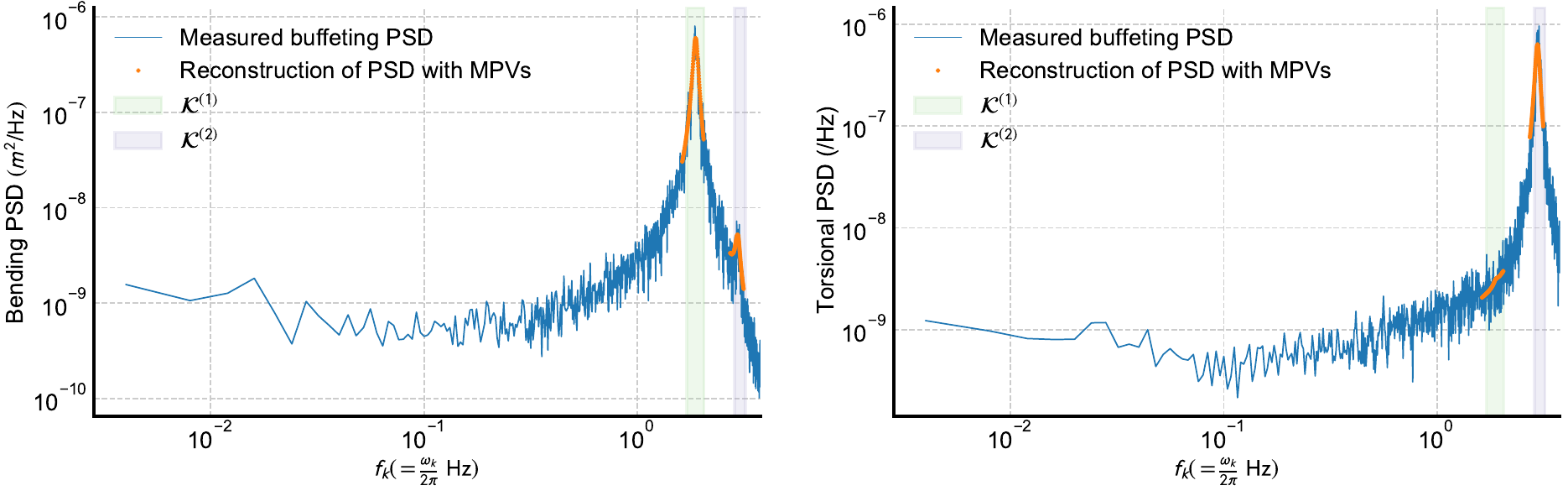}}
    \caption{Case 2: measured buffeting displacement PSDs and reconstruction of PSDs with identified MPVs of FDs and buffeting force PSDs when $U=7.25~\mathrm{m/s}$ ($K_{h}=1.4547,K_{\alpha}=2.3399$)}
	\label{fig:XHM_PSD_reconstruction}
\end{figure}

Fig.~\ref{fig:XHM_PSD_reconstruction} shows the measured buffeting displacement PSDs and the reconstructed PSDs with identified MPVs, where a good consistency is found. In practice, comparison between the measured PSDs and the reconstructed PSDs is of vital importance for the identification process of a general bridge section, which is an efficient method to verify the identified results.

\begin{figure}[H]
    \centering
    \makebox[\textwidth][c]{\includegraphics[]{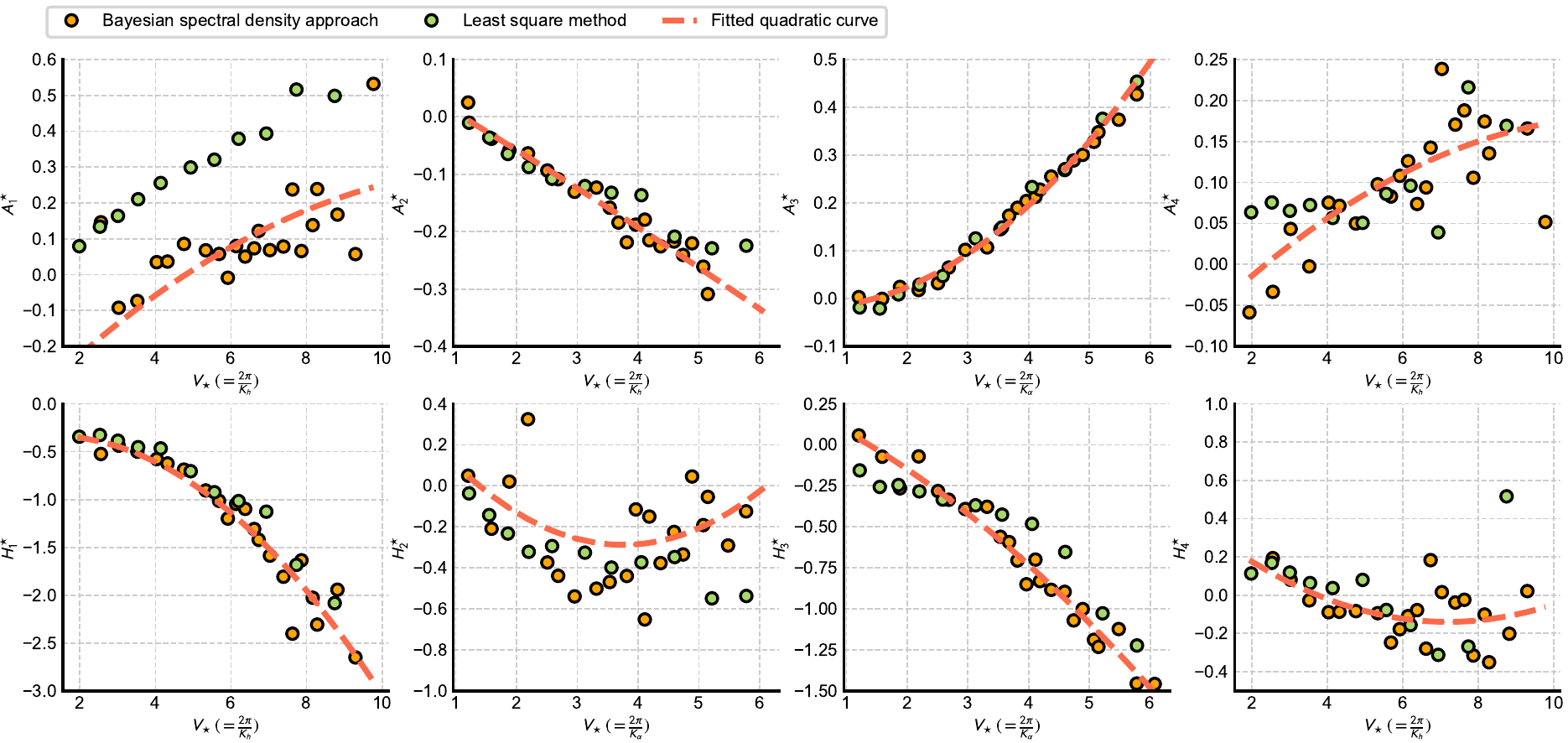}}
    \caption{Case 2: identified MPVs of flutter derivatives of the center-slotted girder model at various reduced velocities by Bayesian spectral density approach in turbulent flow versus least square method in uniform flow}
	\label{fig:simu_MPV_uq_XHM}
\end{figure}

Fig.~\ref{fig:simu_MPV_uq_XHM} shows the identified MPVs of FDs at various reduced velocities by the Bayesian spectral density approach in turbulent flow versus the least square method in uniform flow with free vibration data. The orange dashed lines are quadratic curves fitted by the identified MPVs at various reduced velocities of the Bayesian spectral density approach. Except $A_{1}^{\star}$, the other FDs are consistent when identified by two different methods. The latent reasons for the deviation in $A_{1}^{\star}$ need to be investigated further in the future. Anyway, the bias in $A_{1}^{\star}$ does not affect the reconstruction of buffeting displacement PSDs due to the consistency shown in Fig.~\ref{fig:XHM_PSD_reconstruction}.

\section{Conclusions}
This paper illustrates the application of Bayesian spectral density approach on the identification and uncertainty quantification of flutter derivatives, where the buffeting force PSDs can also be obtained. The identification of flutter derivatives is modeled from a probabilistic perspective based on the complex Wishart distribution. Compared with previous time-domain methods (e.g., least square method and stochastic subspace identification technology), the Bayesian spectral density approach is operated in frequency domain, which has advantages over time-domain methods. In uniform flow using free vibration, for example, when the mean wind speed is high, the coupled damping ratio (combined action from structural damping and aerodynamic damping) will increase fast, which means that the free vibration duration is very short. Short free vibration duration can greatly influence the identification accuracy, which is the reason why the identified results at high wind speed are not reliable by least square method. However, with long enough measuring time, the statistical characteristics of buffeting displacement PSDs in frequency domain are stable and reliable. Because the grids are installed to generate the turbulent flow while conducting the wind tunnel test, the maximum of generated wind speed is relatively lower than that without grids. The advantages of the Bayesian spectral density approach are expected to be more prominent if the mean wind speed is high, which needs to be investigated in the future if a wind tunnel with more wide speed range is available. The identified results in numerical simulations, thin plate model test, and center-slotted girder model test are all satisfactory. In brief, the approach proposed in this paper offers a new angle of view (i.e., probabilistic perspective) for the inference of aerodynamic parameters in wind engineering.

\section*{CRediT authorship contribution statement}

{\textbf{Xiaolei Chu}}: Writing - original draft, Conceptualization, Formal analysis, Investigation, Methodology, Validation, Visualization.
{\textbf{Wei Cui}}: Conceptualization, Supervision, Writing - review $\&$ editing, Funding acquisition.
{\textbf{Peng Liu}}: Methodology.
{\textbf{Lin Zhao}}: Funding acquisition.
{\textbf{Yaojun Ge}}: Supervision, Funding acquisition.

\section*{Acknowledgments}
The authors appreciate PhD candidates Shengyi Xu, Zilong Wang, and Teng Ma for their assistance in the wind tunnel tests. The authors gratefully acknowledge the support of National Natural Science Foundation of China (52008314, 51978527, 52078383). Any opinions, findings and conclusions are those of the authors and do not necessarily reflect the reviews of the above agencies.

\section*{Declaration of competing interest}
The authors declare that they have no known competing financial interests or personal relationships that could have appeared to influence the work reported in this paper.

\bibliography{Bayesian_approach.bib}
\end{document}